\documentclass[prb,twocolumn,showpacs,preprintnumbers,amsmath,amssymb,floatfix]{revtex4}
\usepackage{graphicx}
\usepackage{dcolumn}
\usepackage{bm}

\begin{document}

\title{Thermodynamics of low dimensional spin-1/2 Heisenberg
ferromagnets in an external magnetic field within Green function 
formalism}
\author{T. N. Antsygina, M. I. Poltavskaya, I. I. Poltavsky, and K.
A. Chishko}
\email{chishko@ilt.kharkov.ua}
\affiliation{B. Verkin Institute for Low Temperature Physics and
Engineering, 47 Lenin Ave., 61103 Kharkov, Ukraine}

\date{\today}

\begin{abstract}
The thermodynamics of low dimensional spin-1/2 Heisenberg ferromagnets
(HFM) in an external magnetic field is investigated within a second-order
two-time Green function formalism in the wide temperature and field range.
A crucial point of the proposed scheme is a proper account of the analytical 
properties for the approximate transverse commutator Green function 
obtained as a result of the decoupling procedure. A good quantitative 
description of the correlation functions, magnetization, susceptibility, 
and heat capacity of the HFM on a chain, square and triangular lattices 
is found for both infinite and finite-sized systems. The dependences 
of the thermodynamic functions of 2D HFM on the cluster size are studied.
The obtained results agree well with the corresponding data found by Bethe 
ansatz, exact diagonalization, high temperature series expansions, and 
quantum Monte Carlo simulations.

\end{abstract}


\maketitle

\section {INTRODUCTION}

Quantum Heisenberg model with ferromagnetic exchange is used extensively 
to interpret the thermodynamic and magnetic properties of low dimensional 
(1D and 2D) physical systems. Examples of quasi-one-dimensional ferromagnets 
whose properties can be explained within the Heisenberg model are organic 
p-NPNN compounds \cite{Tak1,Tak2,Tak3,Tak4,Tak5} and cuprates TMCuC.\cite{LW,WL} 
The ferromagnetic insulators such as K$_2$CuF$_4$, Cs$_2$CuF$_4$, La$_2$BaCuO$_5$, 
Rb$_2$CrCl$_4$ \cite{FW1,FW2,MK} and quantum Hall ferromagnets \cite{MAG,BDP,RSach} 
provide examples of the Heisenberg system on a square lattice. A unique example 
of a spin-1/2 magnet on a triangular lattice is $^3$He bilayers adsorbed 
on graphite.\cite{G,Vil,San1,San2,Mor1,Mor2,God2,Rog1} At high coverages 
the second layer proved to be a solid ferromagnet whose thermodynamics can be 
described by HFM with a high degree of accuracy.\cite{Rog1,Gd1,Gd2,GR,A1} 
Nowadays, considerable study is being given to $^3$He monolayers on 
$^4$He-preplated graphite substrates. In these systems under high enough pressure 
a solid $^3$He monolayer with ferromagnetic exchange is formed. \cite{ATR, Saund}

Experimental research of the aforementioned low dimensional magnets is carried out 
intensively, in particular, in the presence of an external magnetic field. 
To interpret the experimental data it is necessary to develop a quantitative 
description  of the HFM thermodynamics at arbitrary  magnetic fields and temperatures.
The two-time Green function formalism is quite appropriate for this purpose. 
The method based on one or another decoupling scheme for higher Green functions 
results in a closed set of self-consistent equations for thermodynamic 
averages.\cite{Zubarev,Rudoy,book,FK} Random phase approximation (RPA) is 
the simplest variant of such scheme with decoupling at the first step.\cite{Tja} 
Being applied to the low dimensional systems it gives satisfactory results
at high fields whereas at low and intermediate fields RPA describes
the thermodynamics only on a qualitative level.

A quantitative description of 1D and 2D spin systems can be obtained within
a more complicated scheme originally proposed in Ref. \onlinecite{KY}  for 1D
HFM in zero magnetic field. The scheme is based on the decoupling of higher
Green functions at the second step with introducing the vertex parameters
to be found. A proper choice of the vertex parameters makes it possible
to retain some relations that must hold true at the exact solving of
the problem. As a result, the theory is built in terms of the correlation
functions and the vertex parameters obeying the self-consistent
set of equations. In Refs. \onlinecite{KY,KY1,ShT,AS1,AS2,A} a single 
vertex parameter was chosen so as to satisfy the sum rule. One vertex 
parameter turned out to be quite enough to describe quantitatively 
the thermodynamics of 1D and 2D (on a square and triangular lattices) 
ferromagnets in zero field.

When employing the above-mentioned scheme to the spin systems in an 
external magnetic field, along with the correlators we have to determine at 
least three additional functions of temperature and field: two vertex 
parameters and magnetization. To do this we need three relations two of which 
are quite evident from the properties of the spin-1/2 operator
\begin{equation}
\langle (S^z)^2 \rangle=\frac{1}{4}, \qquad
\langle S^z \rangle=\frac{1}{2}-\langle S^-S^+\rangle,
\label{eq:sum_rules1}
\end{equation}
where angular brackets denote thermodynamic averaging. The choice of 
the third condition is not so apparent. In Ref. \onlinecite{german1} a 
second-order Green function scheme was applied to HFM chain and HFM on 
a square lattice. As the third condition the authors of Ref. \onlinecite{german1}
used the exact representation of the internal energy through the
transverse Green function.\cite{Tja,book}

The aim of the present work is to calculate the thermodynamic functions 
of HFM on a triangular lattice in an external magnetic field using a
second-order Green function formalism. As compared to a chain and  square 
lattice,\cite{Tkh1D,MK1,RSach,german1,NT} HFM on a triangular lattice in a 
magnetic field is much less investigated. For this case high temperature 
series expansion (HTSE),\cite{Baker} low-temperature asymptotics for the 
magnetization calculated within the spin wave approximation,\cite{Kop,Godfrin1} 
temperature dependences of the magnetization found by quantum Monte Carlo 
simulations (QMC) on a $16\times 16$ cluster,\cite{Kop} and some results 
obtained by the renormalization group technique \cite{Godfrin} (RGT) are 
known. However, none of these approaches gives a complete description of 
the thermodynamics in the whole temperature and field ranges, and a 
second-order Green function method is expected to fill the gap in our knowledge.

In the present work we show that this method is more effective when 
the conditions determining the magnetization and vertex parameters
result from the fundamental principles. Clearly the relations 
(\ref{eq:sum_rules1}) are just of this kind. It is equally important
to retain the analytical properties \cite{Zubarev,book,FK} of the 
Green functions in the approximate approach. Such a requirement 
for the transverse commutator Green function provides a basis for 
the third condition in our theory. Note also that by appropriate choice 
of variables the set of self-consistent equations for the correlators, 
vertex parameters, and magnetization can be written in a universal form 
suited not only for a triangular lattice but also for a chain and square 
lattice. A good agreement of our results obtained for the three types 
of lattices with the corresponding data available from literature confirms 
the efficiency of the used scheme.

In Sec. II, the statement of the problem is formulated and the self-consistent
set of equations for the correlators, magnetization, and vertex parameters 
is derived. In Sec. III, the proposed scheme is applied to HFM on a chain, 
square and triangular lattices. The obtained results are compared with the
corresponding data found within other methods. Some concluding remarks are 
made in Sec. IV.

\section {STATEMENT OF THE PROBLEM}

The Hamiltonian of the system is given by
\begin{equation}
H=-\frac{J}{2}\sum_{\bf f,\mbox{\scriptsize{\boldmath $\delta$}}}
{\bf S}_{\bf f}{\bf S}_{\bf f+\mbox{\scriptsize{\boldmath $\delta$}}}-
h\sum_{\bf f}S_{\bf f}^z,
\label{eq:Hamiltonian}
\end{equation}
where ${\bf S_f}$ is the spin-half operator at site ${\bf f}$,
$\mbox{\boldmath $\delta$}$ is a vector connecting nearest neighbors, 
$J>0$ is an exchange integral, $h=2 \mu B$, $\mu$ is the magnetic moment 
of a particle, $B$ is an external magnetic field.

To calculate spin-spin correlators, it is necessary to find two
retarded commutator single-particle Green functions:
$\langle\langle S^z_{\bf f}|S^z_{{\bf f}^\prime}
\rangle\rangle$, $\langle\langle S^\sigma_{\bf f}|S^{-\sigma}_
{{\bf f}^\prime}\rangle\rangle$\, $(\sigma=\pm)$.
We write down equations of motion for these two functions and make 
the decoupling of the higher Green functions on the second step according 
to the scheme proposed in Ref. \onlinecite{german1}
\begin{eqnarray}
S^{\sigma}_iS^{\sigma}_jS^{-\sigma}_l=\alpha_{\perp}
(\langle S^{\sigma}_jS^{-\sigma}_l\rangle S^{\sigma}_i+
\langle S^{\sigma}_iS^{-\sigma}_l\rangle S^{\sigma}_j),\nonumber
\\
S^{z}_iS^{z}_jS^{\sigma}_l=\alpha_{\perp}
\langle S^{z}_iS^{z}_j\rangle S^{\sigma}_l,
\quad
S^{\sigma}_iS^{-\sigma}_jS^{z}_l=\alpha_{z}
\langle S^{\sigma}_iS^{-\sigma}_j\rangle S^{z}_l,\nonumber
\\
i\neq j \neq l,\qquad i\neq l,\qquad
\label{eq:Decoupling}
\end{eqnarray}
where $\alpha_{\perp}$ and $\alpha_{z}$ are the vertex parameters.

After a number of manipulations we finally obtain for the time-space 
Fourier component
$\langle\langle S^z_{\bf k}|S^z_{-{\bf k}}
\rangle\rangle_\omega$
\begin{equation}
\langle\langle S^z_{\bf k}|S^z_{-{\bf k}}
\rangle\rangle_\omega=\frac{Jc_1\gamma_0}{4\pi}\,\frac{1-\Gamma_{\bf 
k}}
{\omega^2-(\omega^z_{\bf k})^2},
\label{eq:Fourier1}
\end{equation}
where
\begin{equation}
(\omega^z_{\bf k})^2=\frac{J^2\gamma_0}{2}(1-\Gamma_{\bf k})
\left[\Delta_z+\gamma_0\tilde c_1(1-\Gamma_{\bf k})\right],
\label{eq:Om_z}
\end{equation}
\begin{equation}
\Delta_z=1+\tilde c_2-(\gamma_0+1)\tilde c_1.
\label{eq:Del_z}
\end{equation}
Here the following correlation functions have been introduced

\begin{equation}
c_1=2\langle S^{\sigma}_{\bf f}S^{-\sigma}_{\bf 
f+\mbox{\scriptsize{\boldmath $\delta$}}}\rangle, \,\,
c_2=2\sum_{\mbox{\scriptsize{\boldmath $\delta$}} }\,^{\prime}\langle 
S^
{\sigma}_{\bf f+\mbox{\scriptsize{\boldmath $\delta$}}}S^{-\sigma}_
{\bf f+\mbox{\scriptsize{\boldmath $\delta$}}^{\prime}}\rangle,
\,\,
\tilde c_{1,2}=\alpha_{z}c_{1,2}.
\label{eq:c_corr}
\end{equation}
The primed sum indicates that the term with $\mbox{\boldmath 
$\delta$} =
\mbox{\boldmath $\delta$}^{\prime}$ is omitted in it.
The structure factor $\Gamma_{\bf k}$ is defined as
\begin{equation}
\Gamma_{\bf k}=\frac{1}{\gamma_0}
\sum_{\mbox{\scriptsize{\boldmath $\delta$}}}\exp(i{\bf 
k}\mbox{\boldmath $\delta$}),
\label{eq:structure_factor}
\end{equation}
where the coordination number $\gamma_0$ is equal to 2 for a chain, 4 
for a square lattice,
and 6 for a triangular lattice.

Fourier transform $\langle\langle S^\sigma_{\bf k}|S^{-\sigma}_
{-{\bf k}}\rangle\rangle_\omega$ can be written as
\begin{equation}
\langle\langle S^\sigma_{\bf k}|S^{-\sigma}_
{-{\bf k}}\rangle\rangle_\omega=\frac{1}{2\pi}\sum_{l=1,2}
\frac{A^\sigma_{l,{\bf k}}}{\omega-\Omega_{l,{\bf k}}^\sigma}\,.
\label{eq:Fourier2}
\end{equation}
Here
\begin{equation}
\Omega_{l,{\bf k}}^\sigma=h\sigma+(-1)^l\omega^\perp_{\bf k},
\label{eq:Om_l}
\end{equation}
\begin{equation}
(\omega^\perp_{\bf k})^2=\frac{J^2\gamma_0}{2}(1-\Gamma_{\bf k})
\left[\Delta_\perp+\gamma_0\tilde b_1(1-\Gamma_{\bf k})\right],
\label{eq:Om_p}
\end{equation}
\begin{equation}
\Delta_\perp=1+\tilde b_2-(\gamma_0+1)\tilde b_1,
\label{eq:Del_p}
\end{equation}
\begin{equation}
A_{l,{\bf k}}^\sigma=\sigma\langle S^z \rangle + \frac{(-
1)^lJb_1\gamma_0}
{2\omega^\perp_{\bf k}}(1-\Gamma_{\bf k}),
\label{eq:A_l}
\end{equation}
where $\langle S^z \rangle$ is the magnetization.
Due to the presence of the external magnetic field, $\langle S^z \rangle$
is nonzero at any finite temperature.

The correlation functions entering Eqs. (\ref{eq:Om_p})--(\ref{eq:A_l}) are
defined by
\begin{equation}
b_l=\frac{a_l+c_l}{2},  \quad
\tilde b_l=\alpha_\perp b_l, \quad l=1,2,
\label{eq:b_corr}
\end{equation}
\begin{equation}
a_1=4\langle S^z_{\bf f}S^z_{\bf f+\mbox{\scriptsize{\boldmath 
$\delta$}}}\rangle, \qquad
a_2=4\sum_{\mbox{\scriptsize{\boldmath $\delta$}} }\,^{\prime}\langle 
S^
z_{\bf f+\mbox{\scriptsize{\boldmath $\delta$}}}S^z_
{\bf f+\mbox{\scriptsize{\boldmath $\delta$}}^{\prime}}\rangle.
\label{eq:a_corr}
\end{equation}
The Green functions look formally the same for three above-mentioned 
types of lattices. Such universal form has been possible to obtain, 
because instead of the usual correlators describing correlations between
spins which are two steps along the translation vector {\boldmath $\delta$} 
apart, we use linear combinations $c_2$ and $a_2$ defined by Eq. (\ref{eq:c_corr}) 
and Eq. (\ref{eq:a_corr}). The physical meaning of these combinations depends
on the lattice type. For a chain, $c_2$ and $a_2$ are the next nearest neighbor 
correlators. For a square lattice these combinations contain the correlation 
functions between the spin at site $\bf f$ and spins from the second and
third coordination spheres, and for a triangular lattice these combinations 
in addition to the higher order correlators include also $c_1$ and $a_1$. 
For a chain and square lattice Green functions (\ref{eq:Fourier1}) and
(\ref{eq:Fourier2}) coincide with those found in Ref.~\onlinecite{german1}.

Using the spectral relations \cite{Zubarev} we have
 \begin{eqnarray}
a_1=\frac{Jc_1}{N}\sum_{\bf k}\Gamma_{\bf k}g_{\bf k}+
4 \langle S^z\rangle^2,
\nonumber \\
a_2=\frac{Jc_1}{N}\sum_{\bf k} \left(\gamma_0\Gamma_{\bf k}^2-
1\right)g_{\bf k}
+4\left(\gamma_0-1\right)\langle S^z\rangle^2,
\nonumber \\
c_1=\frac{1}{\alpha_{\perp}N}\sum_{\bf k}\Gamma_{\bf k}p_{\bf k},
\nonumber \\
c_2=\frac{1}{\alpha_{\perp}N}\sum_{\bf k}\left(\gamma_0\Gamma_{\bf 
k}^2
-1\right)p_{\bf k},
\label{eq:corr_eq}
\end{eqnarray}
where
\begin{widetext}
\begin{equation}
g_{\bf k}=\frac{\gamma_0}{\omega_{\bf k}^z}\left(1-\Gamma_{\bf 
k}\right)
\coth\left(\frac{\beta\omega_{\bf k}^z}{2}\right), \qquad
p_{\bf k}=
\frac{2\alpha_{\perp}\langle S^z\rangle \sinh\left(\beta h\right)
-J\tilde b_1\gamma_0\left(1-\Gamma_{\bf k}\right)
\sinh\left(\beta\omega_{\bf k}^{\perp}\right)/\omega_{\bf k}^\perp}
{\cosh\left(\beta h\right)-\cosh\left(\beta\omega_{\bf 
k}^{\perp}\right)},
\label{eq:g_p}
\end{equation}
\end{widetext}
$N$ is the total number of sites, $\beta=1/T$. Eq. (\ref{eq:corr_eq}) represents 
the set of equations for the correlation functions $c_l$ and $a_l$. Along with 
these correlators, the set (\ref{eq:corr_eq}) contains the parameters $\alpha_z$, 
$\alpha_{\perp}$, and magnetization $\langle S^z \rangle$ to be also determined.

The vertex parameters are chosen so as to satisfy the sum rules
\begin{equation}
4\langle \left(S_{\bf f}^z\right)^2\rangle=1, \quad
2\langle S_{\bf f}^{\sigma}S_{\bf f}^{-\sigma}\rangle=1+
2\sigma\langle S^z\rangle,
\nonumber
\end{equation}
which using (\ref{eq:g_p}) can be written as
\begin{equation}
\frac{Jc_1}{N}\sum_{\bf k}g_{\bf k}+4\langle S^z\rangle^2=1, \qquad
\alpha_{\perp}=\frac{1}{N}\sum_{\bf k}p_{\bf k}.
\label{eq:sum_rules}
\end{equation}
Finally, in order to close the system (\ref{eq:corr_eq}), (\ref{eq:sum_rules}) 
we need one more equation. It can be found from the following consideration. 
It is known \cite{Zubarev,book} that a commutator Green function must not have 
any pole at $\omega=0$. Clearly Green function (\ref{eq:Fourier1}) does not 
have such pole. A different situation arises with Green function (\ref{eq:Fourier2}). 
When $\omega=0$ its denominator is equal to zero at ${\bf k}={\bf k}_0$ with 
wave vector ${\bf k}_0$ satisfying the equation
\begin{equation}
h=\omega^\perp_{{\bf k}_0}.
\label{eq:k_0}
\end{equation}
Thus, the numerator of Green function (\ref{eq:Fourier2}) must also vanish 
at ${\bf k}={\bf k}_0$, for otherwise this function would have a pole. 
From this condition we get the equation for $\langle S^z \rangle$
\begin{equation}
\langle S^z \rangle = \frac{Jb_1\gamma_0}
{2\omega^\perp_{{\bf k}_0}}(1-\Gamma_{{\bf k}_0}).
\label{eq:S_z_av}
\end{equation}
Note, that in calculating the anticommutator transverse Green function, 
the condition (\ref{eq:S_z_av}) appears automatically without any special 
assumptions (see also Ref. \onlinecite{FK}).

Let us analyze Eq. (\ref{eq:k_0}). The frequency $\omega^{\perp}_{\bf 
k}$ has a maximum $\omega^{\perp}_{max}$ at the edge of the Brillouin zone
\begin{equation}
\left(\omega^{\perp}_{max}\right)^2=\gamma_0J^2\left(
\Delta_{\perp}+2\gamma_0\tilde b_1\right).
\label{eq:Om_p_max}
\end{equation}
Since the parameters $\Delta_{\perp}$ and $\tilde b_1$ in Eq. (\ref{eq:Om_p_max})
are functions of temperature, the frequency $\omega^{\perp}_{max}$ depends on 
temperature as well. It can be shown that $\omega^{\perp}_{max}$ decreases 
monotonically from $\omega^{\perp}_{max}=J\gamma_0$ at $T=0$ to 
$\omega^{\perp}_{max}=J\sqrt{\gamma_0}$ at $T \to \infty$. At $h/J<\sqrt{\gamma_0}$ \,  
Eq. (\ref{eq:k_0}) has a real solution for any temperature.
Substituting it into Eq. (\ref{eq:S_z_av}) we obtain the following 
expression for the magnetization
\begin{equation}
\langle S^z\rangle=\frac{J}{4h\alpha_{\perp}}
\left(\sqrt{\Delta_{\perp}^2+\frac{8h^2\tilde b_1}{J^2}}
-\Delta_{\perp}\right)\,.
\label{eq:S_z}
\end{equation}
In the field range $\sqrt{\gamma_0}<h/J<\gamma_0$ the real solution 
of Eq. (\ref{eq:k_0}) exists only at $T<T_0$ (where $T_0$ obeys the 
equation $\omega^{\perp}_{max}(T_0)=h$). Finally, if $h/J>\gamma_0$ 
Eq. (\ref{eq:k_0}) has no real solutions at any temperature.

It is natural to suppose that the expression (\ref{eq:S_z}) for the 
magnetization is valid at arbitrary $h$ and $T$. This assumption provides 
continuity of $\langle S^z\rangle$ as a function of field and temperature.
Eq. (\ref{eq:S_z}) gives correct values of the magnetization at low and high 
fields for arbitrary temperatures and at $T=0$ for arbitrary fields. However, 
the most important thing is that Eq. (\ref{eq:S_z}) provides correct analytical 
properties of the commutator Green function (\ref{eq:Fourier2}) obtained within 
the approximate scheme.

As a result, Eqs. (\ref{eq:corr_eq}), (\ref{eq:sum_rules}), and (\ref{eq:S_z}) 
represent a closed set of seven self-consistent equations for $a_1$, $a_2$, 
$c_1$, $c_2$, $\alpha_{z}$, $\alpha_{\perp}$, and $\langle S^z\rangle$. This set 
can be reduced to three equations for $\tilde b_1$, $\Delta_z$, and $\Delta_{\perp}$
\begin{eqnarray}
1=\frac{Jc_1}{N}\sum_{\bf k}g_{\bf k}+4\langle S^z\rangle^2,
\nonumber \\
2\tilde b_1=\alpha_{\perp}c_1\left[1-\frac{J}{N}\sum_{\bf k}
\left(1-\Gamma_{\bf k}\right)g_{\bf k}\right]+\alpha_{\perp},
\nonumber \\
\Delta_{\perp}=1-\alpha_{\perp}
+\frac{\alpha_{\perp}}{2\alpha_{z}}\left(\Delta_{z}-1\right)+
\nonumber \\
\frac{J\alpha_{\perp}c_1}{2N}\sum_{\bf k}
\left(1-\Gamma_{\bf k}\right)\left(1-\gamma_0\Gamma_{\bf k}\right)
g_{\bf k}.
\label{eq:eq_system}
\end{eqnarray}
The values $c_1$, $\alpha_{\perp}$, and $\langle S^z\rangle$ can be expressed through 
$\tilde b_1$, $\Delta_z$ and $\Delta_{\perp}$ according to (\ref{eq:corr_eq}), 
(\ref{eq:sum_rules}) and (\ref{eq:S_z}). For $\alpha_{z}$ with the help of 
Eqs. (\ref{eq:Del_z}) and (\ref{eq:c_corr}) we have
\begin{equation}
\alpha_{z}=\frac{1-\Delta_{z}}{\left(\gamma_0+1\right)c_1-c_2}.
\label{eq:alpha_z}
\end{equation}
It is easy to see that the replacement $h \to -h$ changes the sign of the magnetization 
and does not change the correlation functions and vertex parameters. Owing to 
condition (\ref{eq:S_z}), Eq. (\ref{eq:g_p}) and thereby Eqs. (\ref{eq:corr_eq}), 
(\ref{eq:sum_rules}) have no singularities. At $h=0$ the system (\ref{eq:eq_system}) 
reduces to that found in Refs. \onlinecite{KY,A}.

The internal energy $E$ per site is given by
\begin{eqnarray}
E=-\frac{J\gamma_0}{8} (2c_1+a_1)-h\langle S^z \rangle.
\label{eq:term_par}
\end{eqnarray}

The efficiency of the proposed scheme can be estimated, first, by comparing
the obtained results with available from literature data found by alternative 
methods and, second, with the help of inherent criteria existing within the 
developed scheme itself. The first criterion implies that at $T \to \infty$ 
the entropy (per site) of the system with spin 1/2 should tend to 
$S(\infty)=\ln 2\approx 0.693$.

The second criterion follows from the relation \cite{book} connecting
the internal energy (\ref{eq:term_par}) and the Green function
$\langle\!\langle S^+_{\bf k}|S^-_{-{\bf k}}\rangle\!\rangle
_\omega$, that can be written as
\begin{eqnarray}
\frac{J\gamma_0}{8}(2c_1+a_1-1)+h\left (\langle S^z\rangle-
\frac{1}{2}\right)-
\nonumber \\
\frac{1}{N}\sum_{\bf k}\int\limits_{-\infty}^{\infty}
d\omega\frac{\left(\varepsilon_{\bf k}+\omega\right){\rm Im}
\langle\!\langle S^+_{\bf k}|S^-_{-{\bf 
k}}\rangle\!\rangle_\omega}{e^{\beta\omega}-1}=0,
\label{eq:sec_crit}
\end{eqnarray}
where $\varepsilon_{\bf_k}=J\gamma_0(1-\Gamma_{\bf k})/2+h$.
The relation (\ref {eq:sec_crit}) becomes the identity with the exact 
Green function and correlators. This is not the case for the Green function 
and correlators found as a result of the decoupling procedure. Dividing 
Eq. (\ref {eq:sec_crit}) by $J\gamma_0\langle S^z\rangle/4$ and
substituting (\ref{eq:Fourier2}) in its left hand side, we get
\begin{widetext}
\begin{eqnarray}
1-\frac{1}{J\gamma_0\langle S^z\rangle N}
\sum_{\bf k}\frac{Jb_1\gamma_0\left(1-\Gamma_{\bf k}\right)
\sinh\left(\beta h\right)-2\langle S^z\rangle \omega^{\perp}_{\bf k}
\sinh\left(\beta \omega^{\perp}_{\bf k}\right)}
{\cosh\left(\beta h\right)-\cosh\left(\beta\omega_{\bf 
k}^{\perp}\right)}
\equiv R
\label{eq:crit_A}
\end{eqnarray}
\end{widetext}
The quantity $R$ is a function of field and temperature.
Due to Eq. (\ref{eq:S_z}) expression (\ref {eq:crit_A}) is singularity-free. 
It is evident that the closer $R$ to zero, the better the approximation. 
Thus, the condition $|R|\ll 1$ can serve as another criterion of the 
approximation efficiency. Below, in discussing the results we will calculate 
$R(h,T)$ and check the fulfillment of this criterion for the proposed scheme.

Note, that employing a similar approach for HFM on a chain and square lattice
(below we refer to it as Green function approximate method (GFAM)) the 
authors of Ref. \onlinecite{german1} instead of Eq. (\ref{eq:S_z}) used the 
condition $R=0$ as one of the equations in the self-consistent set of equations. 
In the following we will compare the thermodynamic functions calculated within 
our scheme with the results found in Ref.~\onlinecite{german1}.

\section {RESULTS AND DISCUSSION}

In the general case the set of equations (\ref{eq:eq_system}) can be solved 
only numerically. In limiting cases analytical results could be obtained. 
At $T=0$ the system (\ref{eq:eq_system}) gives correct values for the sought 
quantities
\begin{equation}
\langle S^z \rangle=\frac{1}{2}\,, \quad c_{1,2}=0, \quad
a_1=1, \quad a_2=\gamma_0-1.
\label{eq: func_T0}
\end{equation}
The same solution is also true for finite temperatures at $h \to \infty$.

In the high temperature limit ($J,h \ll T$) the system (\ref{eq:eq_system}) can
be solved by expanding in $1/T$. Restricting our consideration to the second 
order in $x=J/(4T)$ and $y=h/(2T)$ we obtain the following asymptotic expressions
\begin{eqnarray}
c_1^{as}= x+\frac{x^2}{4} (\gamma_0^2-6\gamma_0+4),
\nonumber \\
c_2^{as}= \frac{x}{4}\left[(\gamma_0-2)(\gamma_0-4)-
x(\gamma_0^2-14\gamma_0+20)\right ],
\nonumber \\
a_1^{as}=c_1^{as}+y^2, \qquad
a_2^{as}=c_2^{as}+(\gamma_0-1)y^2,
\nonumber \\
\langle S^z \rangle^{as} = \frac{y}{2}\left(1+\gamma_0 x \right )\,,
\qquad
\alpha_z^{as}=\alpha_\perp^{as}=1-\frac{x}{3}\,.
\label{eq:assimpt}
\end{eqnarray}
For the heat capacity $C=dE/dT$ to the third order in $x$ and $y$ we get
\begin{eqnarray}
C^{as}=\frac{3\gamma_0 x^2}{2}\left[1+ \frac{x}{2}\left( \gamma_0^2-
6\gamma_0+4\right )\right] +
\nonumber \\
y^2\left(1+3\gamma_0 x\right).
\label{eq:Cas}
\end{eqnarray}

The expressions for the magnetization and heat capacity coincide with those 
obtained by the direct high temperature series expansion.\cite{Baker}
As it follows from Eq.~(\ref{eq:assimpt}) the field-dependent terms 
in expansions for $c_l$ occur in the fourth or higher order in $1/T$.

We will demonstrate the efficiency of the proposed scheme applying it 
to the 1D HFM and HFM on a square lattice. The main attention will be 
paid to low fields, because it is this region that is the most difficult 
for the adequate description within approximate methods.

\subsection{\bf One-dimensional Heisenberg model}

\begin{figure}
\includegraphics[width=0.45\textwidth, trim=0 10 0 0]{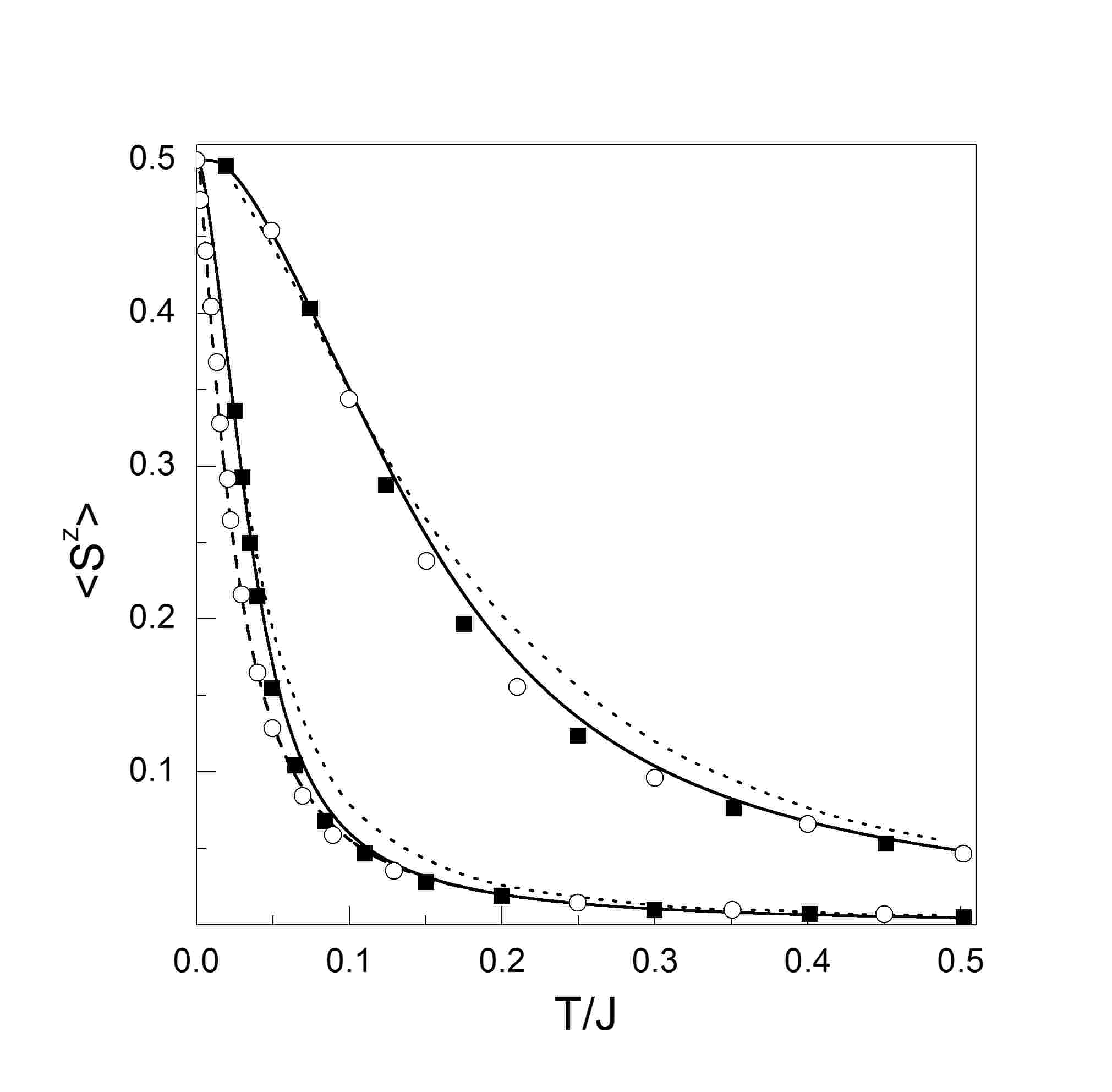}
\caption{\label{fig:epsart1}
Temperature dependences of the magnetization for 1D HFM
at $h/J=$ 0.005 and 0.05 (from left to right).
The infinite system: present theory (solid),
BA\cite{german1} ($\blacksquare$) and GFAM\cite{german1} (dotted).
The cluster: present theory (dashed) and ED\cite{german1}
($\circ$).}
\end{figure}

\begin{figure}
\includegraphics[width=0.45\textwidth, trim=0 10 0 0]{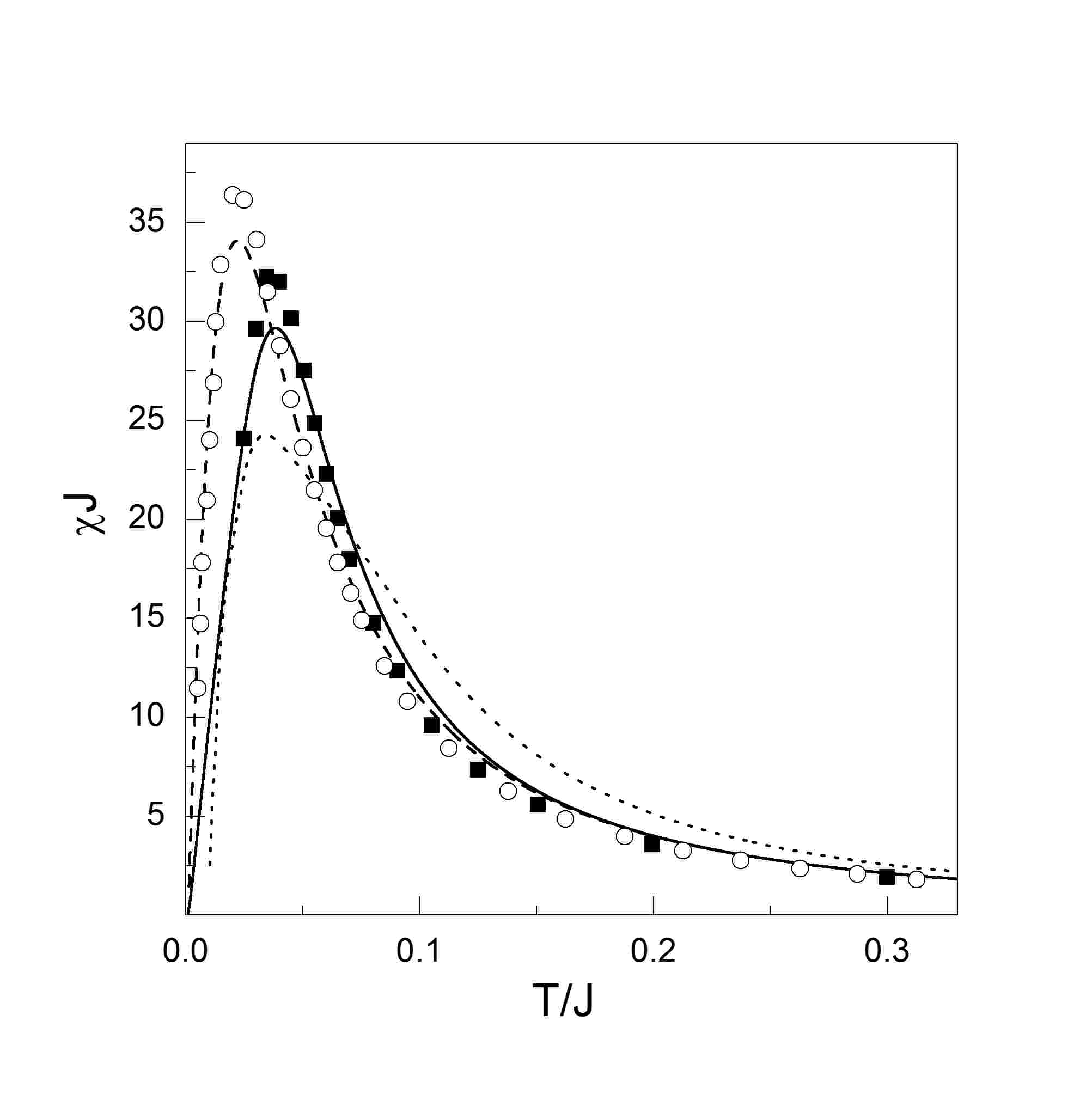}
\caption{\label{fig:epsart2}
Temperature dependences of the susceptibility for 1D HFM
at $h/J=$ 0.005.
The infinite system: present theory (solid),
BA\cite{german1} ($\blacksquare$) and GFAM\cite{german1} (dotted).
The  cluster: present theory (dashed) and ED\cite{german1} ($\circ$).}
\end{figure}

\begin{figure}
\includegraphics[width=0.45\textwidth, trim=0 10 0 0]{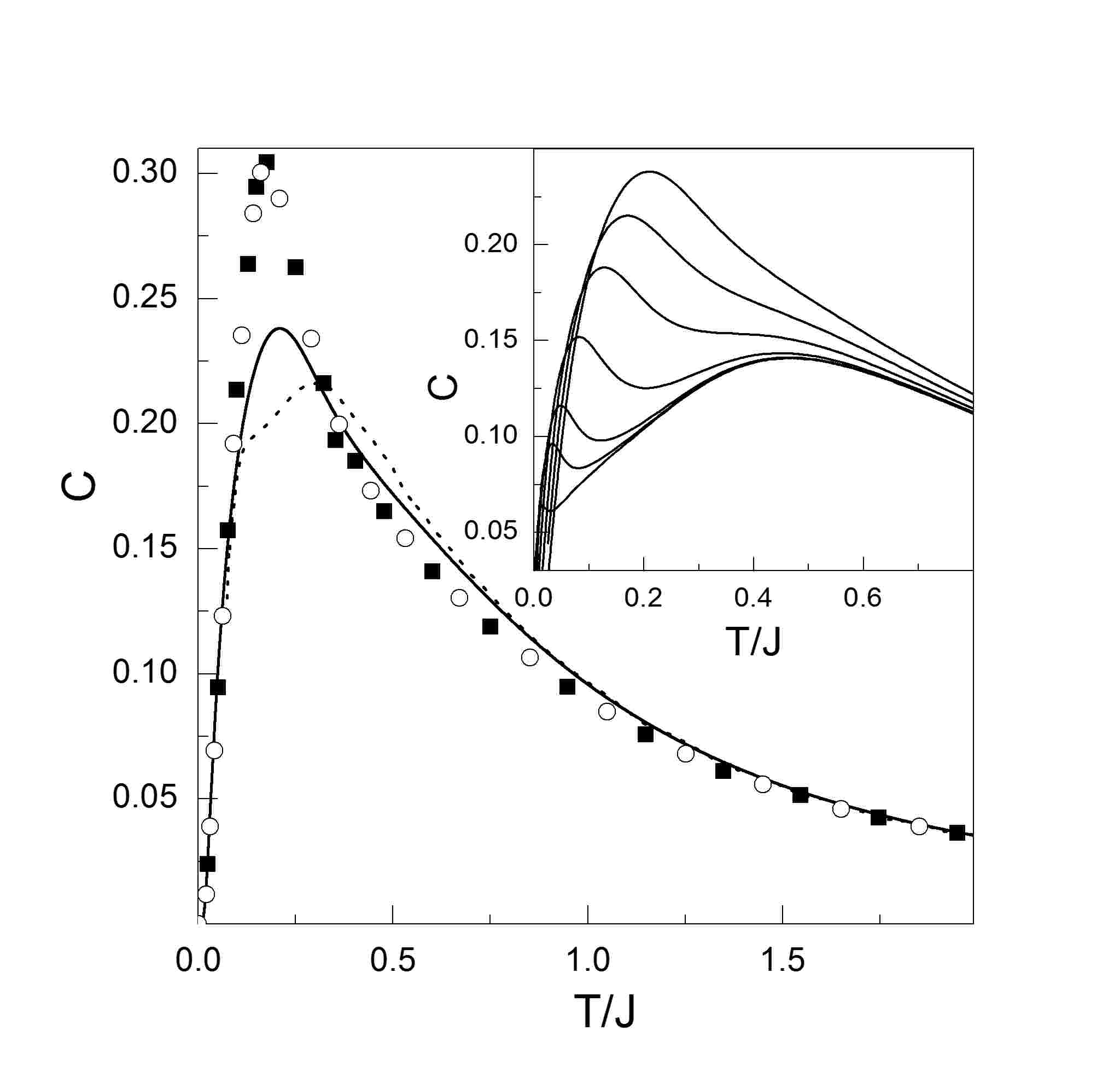}
\caption{\label{fig:epsart3}
Temperature dependences of the heat capacity for 1D HFM at $h/J=$ 0.1.
The infinite system: present theory (solid), BA\cite{german1} ($\blacksquare$), 
GFAM\cite{german1} (dotted). The cluster: ED\cite{german1} ($\circ$).
The inset shows $C(T)$ at low fields $h/J$=0.001, 0.005, 0.01, 0.025, 
0.05, 0.075, 0.1 (from bottom to top).}
\end{figure}

In this subsection we consider 1D HFM. Figs. \ref{fig:epsart1}--\ref{fig:epsart3} 
demonstrate the magnetization, susceptibility $\chi=\partial \langle S^z \rangle 
/\partial h$, and heat capacity vs temperature at low fields obtained within 
our approach. In calculating the thermodynamic functions on clusters, we use 
the periodic boundary conditions. The corresponding dependences found in 
Ref. \onlinecite{german1} by Bethe ansatz (BA) and GFAM for an infinite chain and 
by the exact diagonalization (ED) for a cluster of 16 sites are also shown for 
comparison.

For the magnetization and susceptibility our method yields a good agreement
with the exact results in the whole temperature range. The positions of the 
maxima in our curves for $\chi(T)$ coincide perfectly with those found within 
the exact methods and only a small difference in the peak heights is observed.

At low fields the exact methods indicate the dependence of the thermodynamic
functions on the chain length. As it is seen from Figs. \ref{fig:epsart1}, 
\ref{fig:epsart2}, at $h/J=0.005$ the curves $\langle S^z(T) \rangle$
and $\chi(T)$ obtained by ED for the finite-sized chain differ substantially 
from those found by BA for the infinite system. Our method gives a proper 
description of this effect. At higher fields (Figs. \ref{fig:epsart1}, 
\ref{fig:epsart3}) where the ED and BA results coincide our dependences 
for $N=16$ and $N\to \infty$ also coincide and show a good fit to the exact data.

As it can be seen from Fig. \ref{fig:epsart3} there is a certain disagreement 
between the heat capacities obtained within the exact and approximate methods at 
low fields. This result is quite understandable. Indeed, a similar scheme\cite{KY} 
applied to 1D HFM at $h=0$  gives in the low temperature region sufficiently 
different run of the heat capacity than the exact solution. Nevertheless, even at
$h/J=0.1$ our theory not only gives the correct position of the maximum but
also reproduces a specific bend in the curve $C(T)$ at $T/J \sim 0.3$.
The agreement between $C(T)$ calculated within our scheme and found by the exact 
methods becomes better with increase in field. The inset in Fig. 3 illustrates 
a double-peak structure of the heat capacity that within our method is identified 
at $0<h/J<0.045$. A similar structure of $C(T)$ at low fields was first obtained in
Ref. \onlinecite{german1}.

We calculated the entropy for 1D HFM. It tuned out that the higher is 
the field the closer is the limiting value $S(\infty)$ to $\ln 2$.
For example, at $h/J=0.05$ the entropy is $S(\infty)\simeq 0.631$ and 
at $h/J=1$ it is $S(\infty)\simeq 0.687$. The quantity $R(h,T)$ was also 
found. At low and high temperatures it is practically equal to zero, so 
that the Green function (\ref{eq:Fourier2}) and correlators calculated 
within our method may be considered as satisfying Eq. (\ref{eq:sec_crit}). 
At a given field in the intermediate temperature range where the correlators 
vary rapidly the quantity $R$ is at maximum. At $h/J=0.05$ the maximum 
value of $R$ is $\sim 0.027$ whereas at $h/J=1$ it does not exceed 0.006.
With increase in $h/J$ the quantity $R$ decreases and the difference 
between results obtained by the exact and approximate methods vanishes.

\subsection{\bf HFM on a square lattice}

\begin{figure}
\includegraphics[width=0.45\textwidth, trim=0 10 0 0]{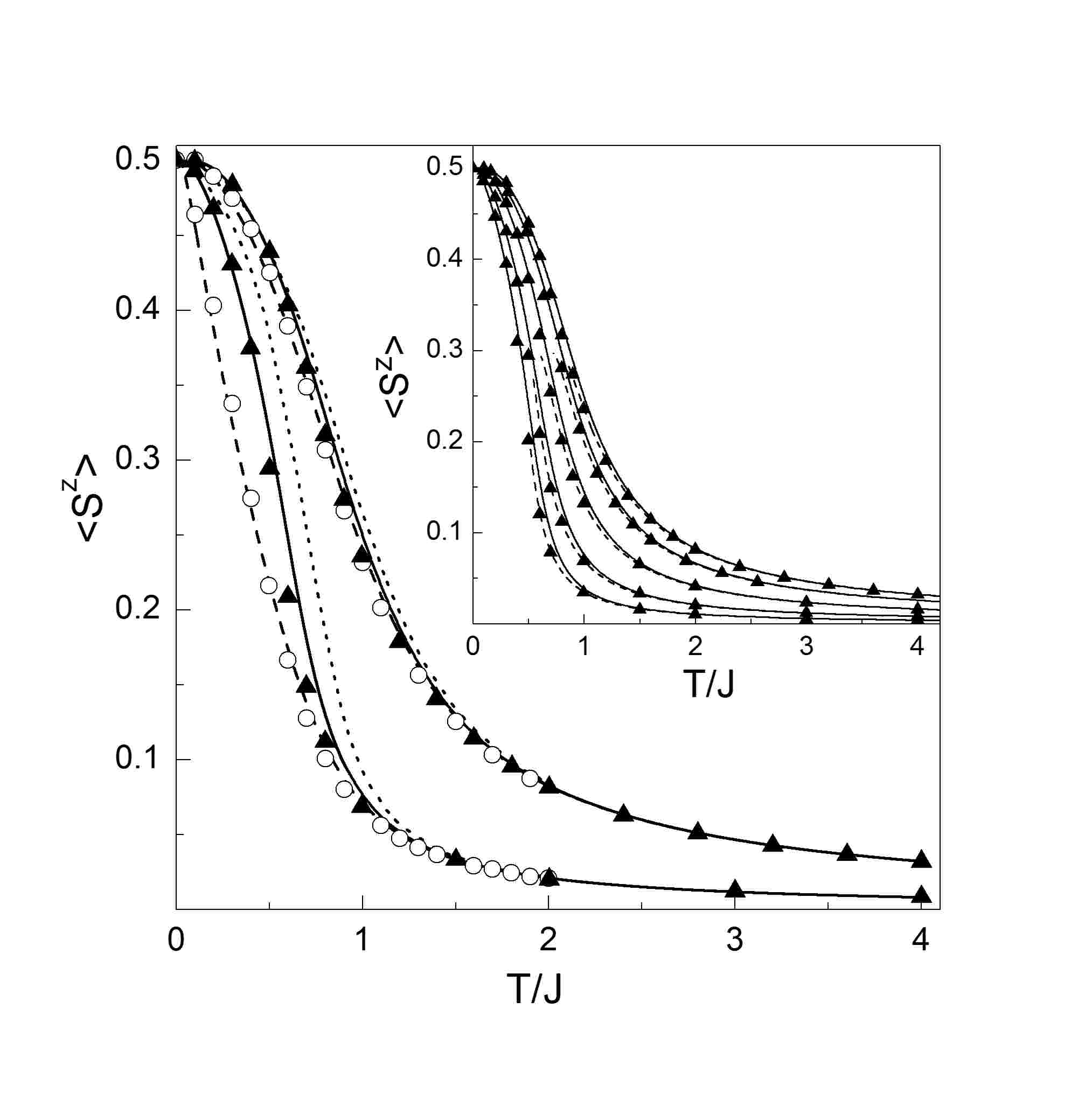}
\caption{\label{fig:epsart4}
Temperature dependences of the magnetization for HFM on
a square lattice at $h/J=$ 0.1 and 0.4 (from left to right).
The infinite system: present theory (solid),
QMC\cite{MK1} ($\blacktriangle$) and GFAM\cite{german1}
(dotted). The $4\times 4$ cluster: present theory (dashed) 
and ED\cite{german1} ($\circ$). The inset shows 
$\langle S^z\rangle$ vs $T/J$ at $h/J$=0.05, 0.1, 0.2, 0.32,
0.4 (solid) in comparison with QMC\cite{MK1} ($\blacktriangle$) 
and HTSE\cite{Baker} (dashed) (from left to right).}
\end{figure}
\begin{figure}
\includegraphics[width=0.45\textwidth, trim=0 10 0 0]{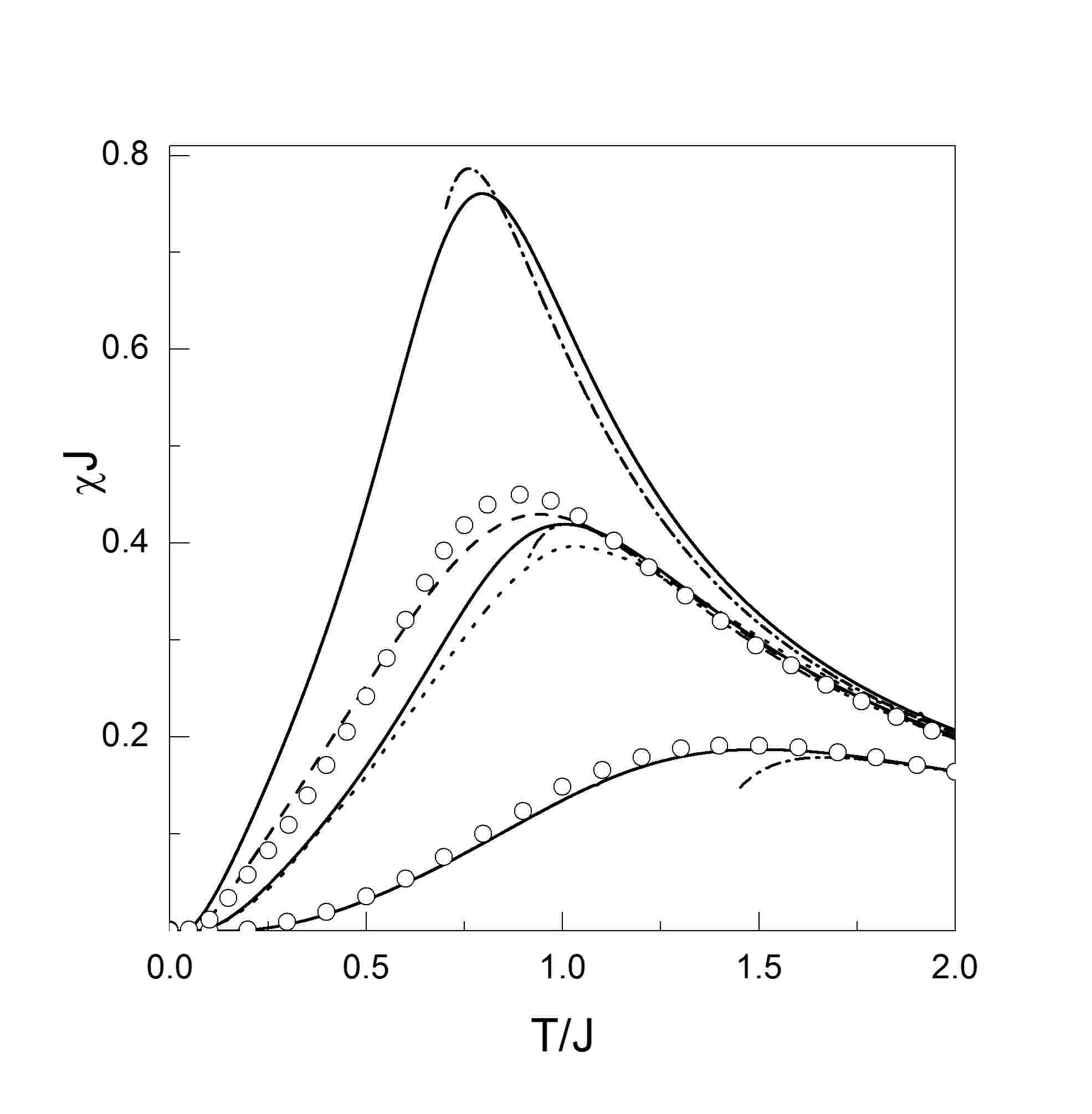}
\caption{\label{fig:epsart5}
Temperature dependences of the susceptibility for HFM on a square
lattice at $h/J=$ 0.2, 0.4, 1.0 (from top to bottom).
The infinite system: present theory (solid), HTSE\cite{Baker} 
(dash-dotted), and GFAM\cite{german1} (dotted).
The $4\times 4$ cluster: present theory (dashed) and ED\cite{german1} 
($\circ$).}
\end{figure}
\begin{figure}
\includegraphics[width=0.45\textwidth, trim=0 10 0 0]{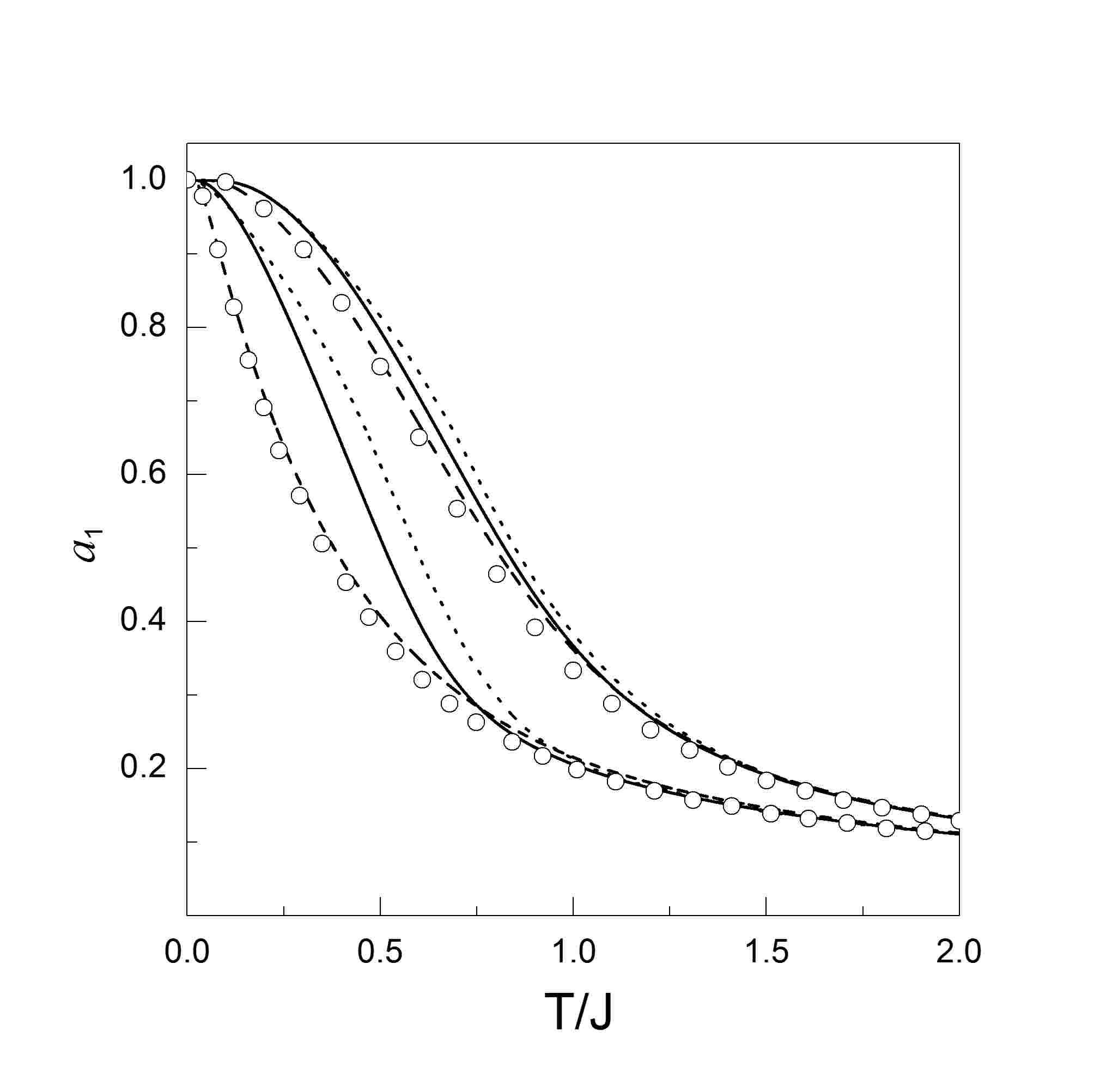}
\includegraphics[width=0.45\textwidth, trim=0 10 0 0]{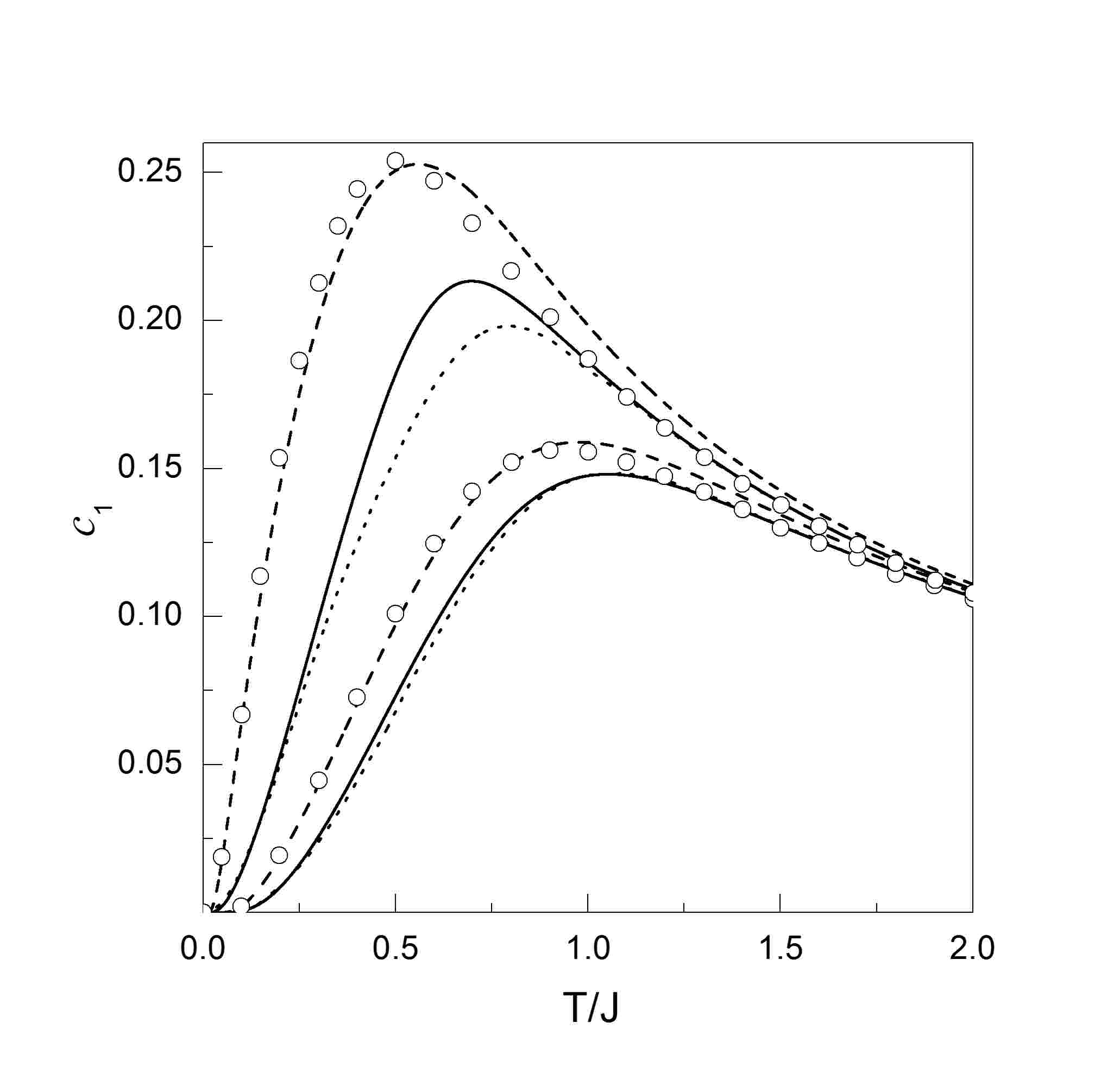}
\caption{\label{fig:epsart6}
Correlation functions $a_1$ (a) and $c_1$ (b) for HFM on
a square lattice at $h/J$=0.1 and 0.4 (from left to right).
The infinite system: present theory (solid) and GFAM\cite{german1} (dotted).
The $4\times 4$ cluster: present theory (dashed) and ED\cite{german1} ($\circ$).}
\end{figure}

Now we proceed to the HFM on a square lattice. Fig. \ref{fig:epsart4} 
demonstrates the magnetization as a function of temperature at 
$h/J=$0.1 and 0.4 for $4\times 4$ and $32\times 32$ square lattice 
clusters together with QMC,\cite{MK1} ED,\cite{german1} and GFAM\cite{german1} 
data. The inset shows the magnetizations for the $32\times 32$ cluster at 
$h/J=0.05, 0.1, 0.2, 0.32, 0.4$ found within the present method, QMC,\cite{MK1} 
and HTSE.\cite{Baker} The comparison between our results and HTSE is possible, 
because at these values of field the magnetizations for the $32\times 32$ cluster 
proved\cite{MK1} to be identical to those for the infinite lattice.
It is seen that the temperature dependences of $\langle S^z \rangle$ 
obtained within our approach are in good agreement with the exact results. 
The proposed scheme reproduces correctly the dependence of $\langle S^z \rangle$ 
on the size of the system as well. The difference between the present results 
and GFAM in Fig. \ref{fig:epsart4} (see also Figs. \ref{fig:epsart1}--\ref{fig:epsart3}) 
testifies that Eq. (\ref{eq:S_z}) is more preferable as compared to the condition $R=0$ 
for a quantitative description of the low dimensional HFM at small fields. It is also 
evident, that the thermodynamic functions of the square lattice HFM are more sensitive 
to the choice of the condition for $\langle S^z \rangle$ than the thermodynamic functions 
for 1D HFM. Naturally, such a choice is expected to be even more critical for the 
lattices with larger coordination numbers (for example, a triangular lattice).

Fig. \ref{fig:epsart5} illustrates the susceptibility $\chi(T)$ for the square 
lattice together with $\chi(T)$ found by HTSE,\cite{Baker} ED,\cite{german1} 
and GFAM.\cite{german1}

The temperature dependences of the correlation functions $a_1$ and $c_1$
at $h/J=0.1$ and $0.4$ calculated for the $4\times 4$ cluster as well 
as for the infinite lattice are presented in Fig. \ref{fig:epsart6}. 
The ED and GFAM results are added for comparison. At low fields a clearly 
defined dependence on the size of the system is seen. Our results for the 
infinite lattice differ noticeably from GFAM. For the $4\times 4$ cluster 
a good agreement with the dependences calculated by ED is observed.

The limiting value of entropy for the HFM on the square lattice is 
$S(\infty)=0.651$ at $h/J=0.05$ and $S(\infty)=0.684$ at $h/J=1$, 
which is very close to the exact value $\ln 2$. The maximum value of $R$ 
is $\simeq 0.058$ at $h/J=0.05$ and $\simeq 0.014$ at $h/J=1$.

Thus, the results of subsections A, B show, that the theory based on 
the correct accounting for the analytical properties of Green functions
gives an adequate description of the thermodynamic functions for the 
systems under consideration in the wide field and temperature range.

\subsection{\bf HFM on a triangular lattice}

\begin{figure}
\includegraphics[width=0.45\textwidth, trim=0 10 0 0]{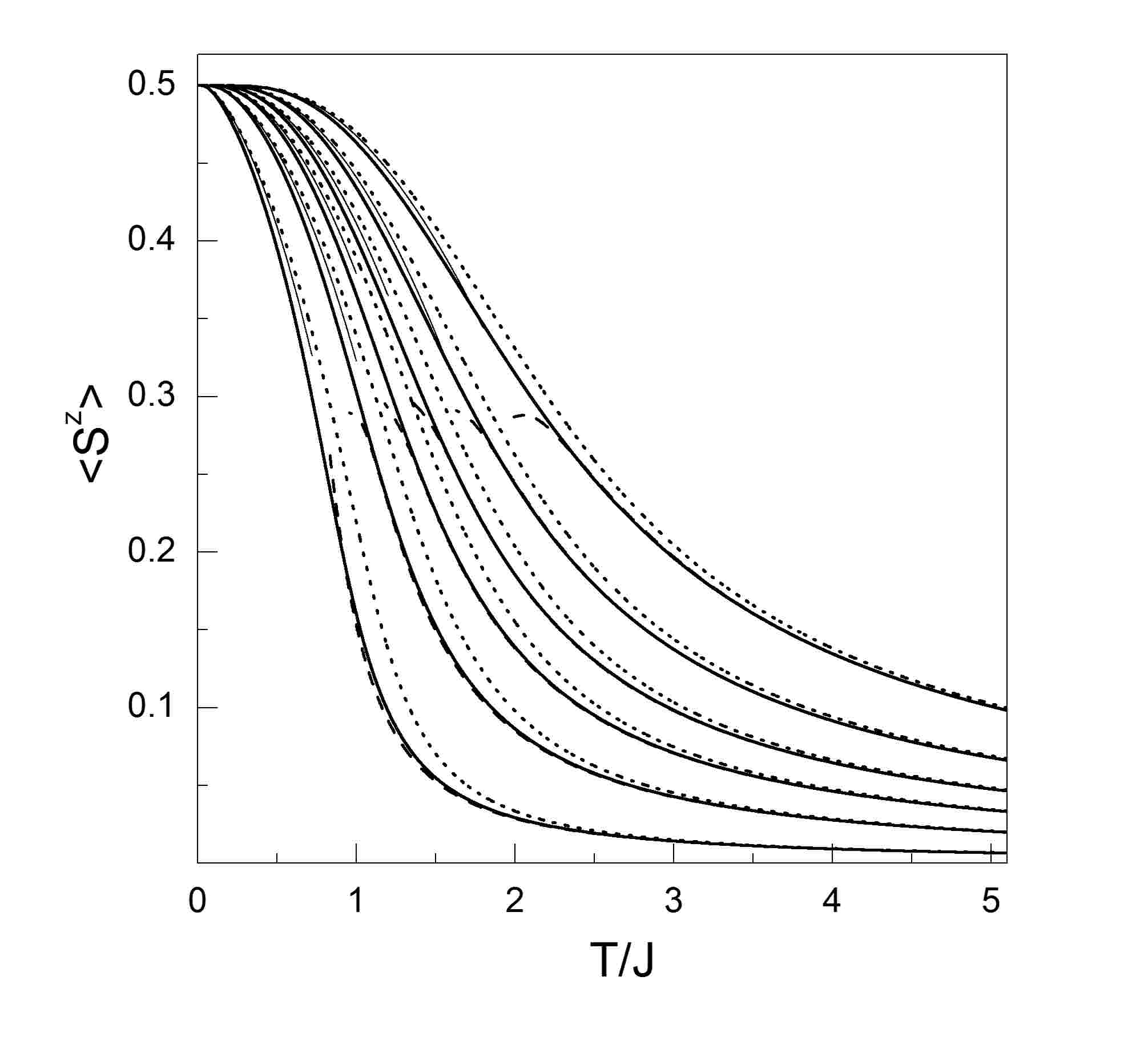}
\caption{\label{fig:epsart7}
Temperature dependences of the magnetization for HFM on a triangular
lattice at $h/J=$ 0.1, 0.3, 0.5, 0.7, 1.0, 1.5 (from left to right).
The present theory (solid), HTSE\cite{Baker} (dashed),
RGT (thin lines), and RPA (dotted).}
\end{figure}

In this subsection we consider HFM on a triangular lattice in an external
magnetic field with peculiar attention concentrated on small and intermediate 
fields. Fig. \ref{fig:epsart7} represents the temperature dependences of the  
magnetization at different values of $h/J$. It is seen that our results 
agree closely with HTSE\cite{Baker} up to the point of HTSE applicability, 
whereas the RPA curves coincide with HTSE only at relatively high temperatures.
In the intermediate temperature range the RPA results differ sufficiently 
from ours even at $h/J$=1.5 reproducing the temperature behavior of 
$\langle S^z\rangle$ only qualitatively. We have also calculated the 
magnetization at low temperatures using the renormalization group technique.
The results obtained by both our approaches are in good agreement.

\begin{figure}
\includegraphics[width=0.45\textwidth, trim=0 10 0 0]{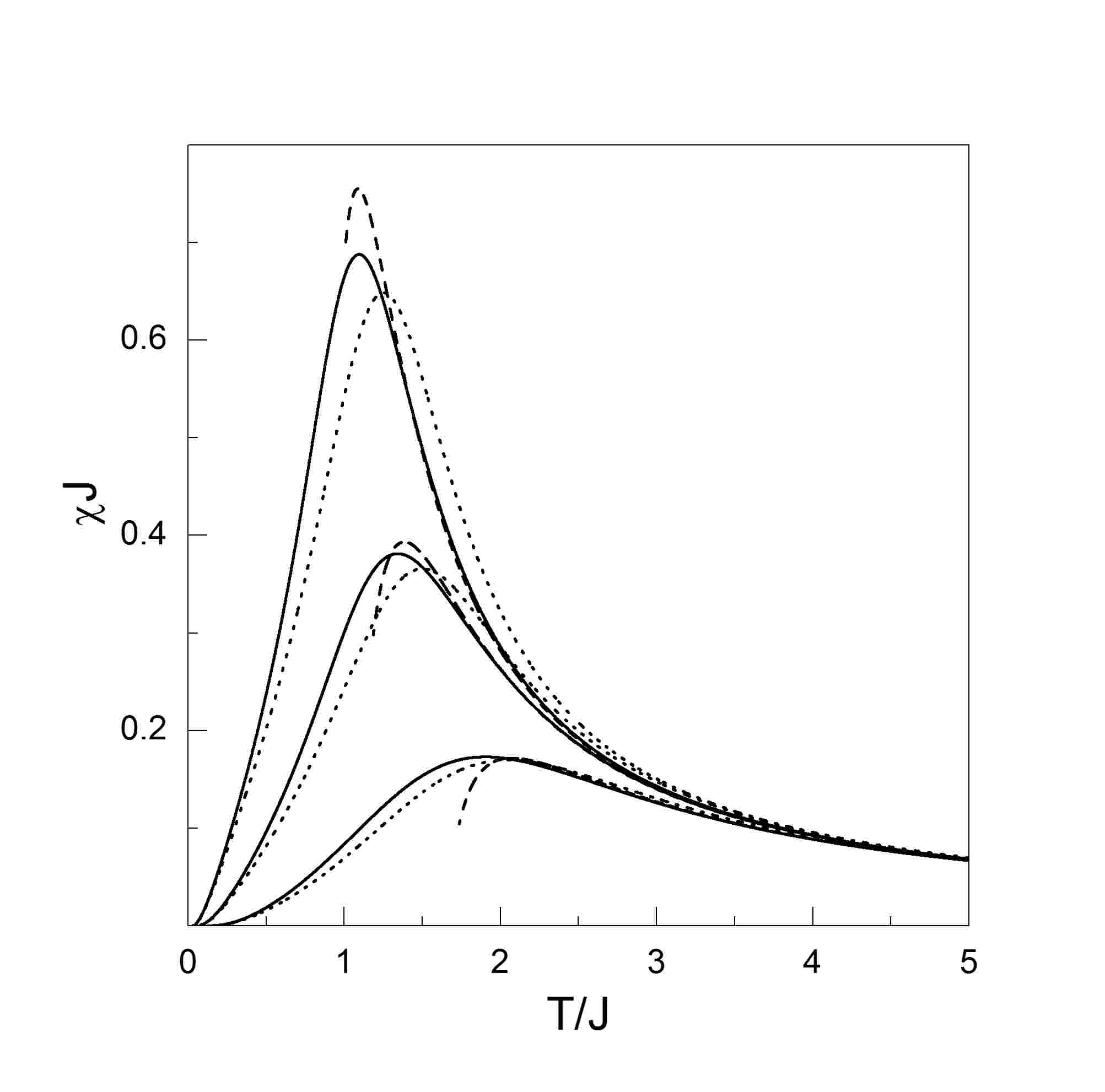}
\caption{\label{fig:epsart8}
Temperature dependences of the susceptibility for HFM on a triangular
lattice at $h/J=$ 0.2, 0.4, 1.0 (from top to bottom).
The present theory (solid), HTSE\cite{Baker} (dashed), and RPA (dotted).}
\end{figure}
Fig. \ref{fig:epsart8} illustrates the temperature behavior of the 
susceptibility. Analysis shows that with increase in field the maximum 
in $\chi(T)$ decreases and shifts to higher temperatures. At $h/J\geq0.1$ 
the height of the maximum as a function of $h/J$ with a great degree of 
accuracy is described by a power law
\begin{equation}
\chi_{max} = a\left(\frac{h}{J}\right)^b, \quad  a=0.1696, \quad
b=-0.8634.
\label{eq:chi_max}
\end{equation}

\begin{figure}
\includegraphics[width=0.45\textwidth, trim=0 10 0 0]{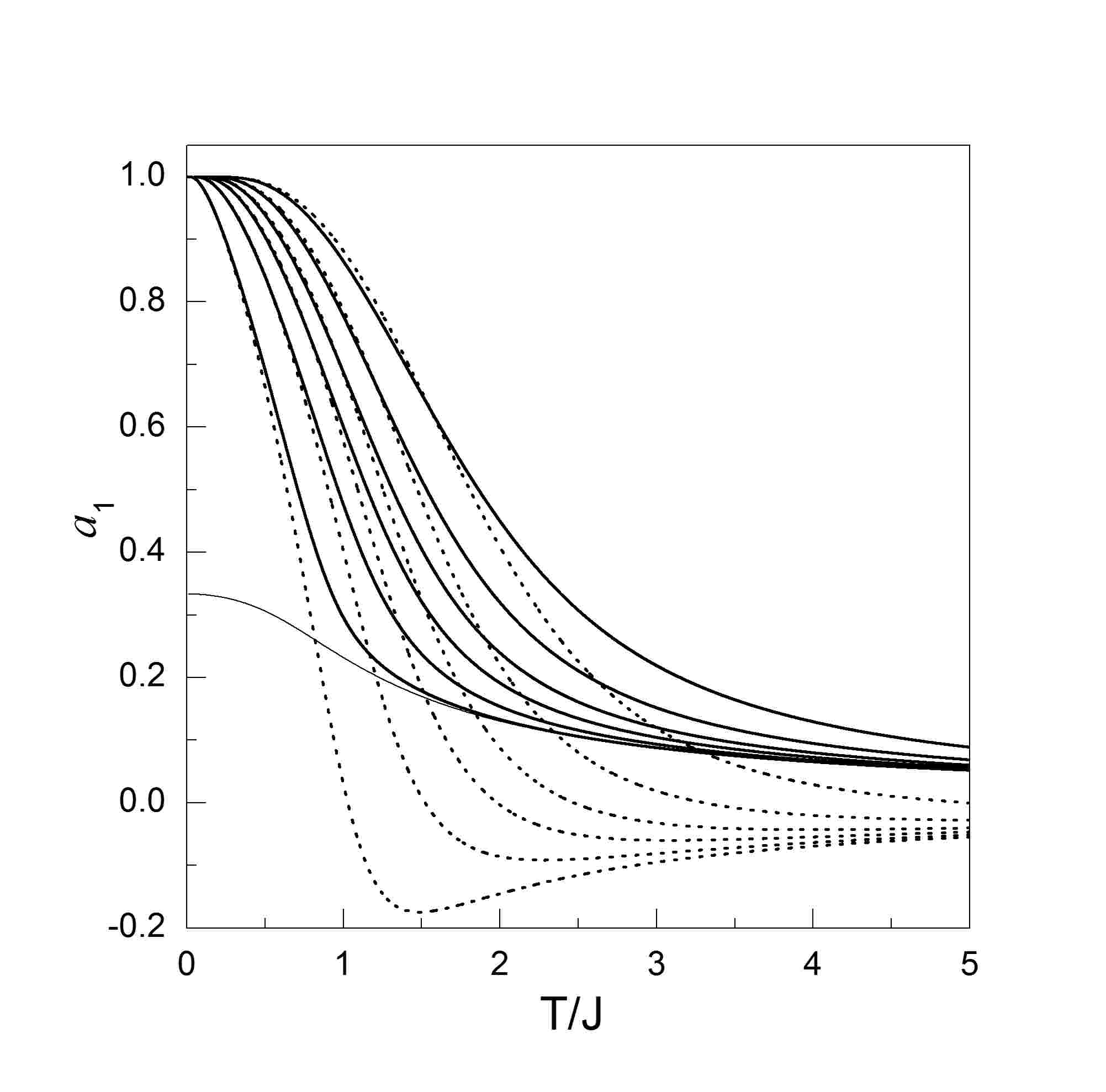}
\includegraphics[width=0.45\textwidth, trim=0 10 0 0]{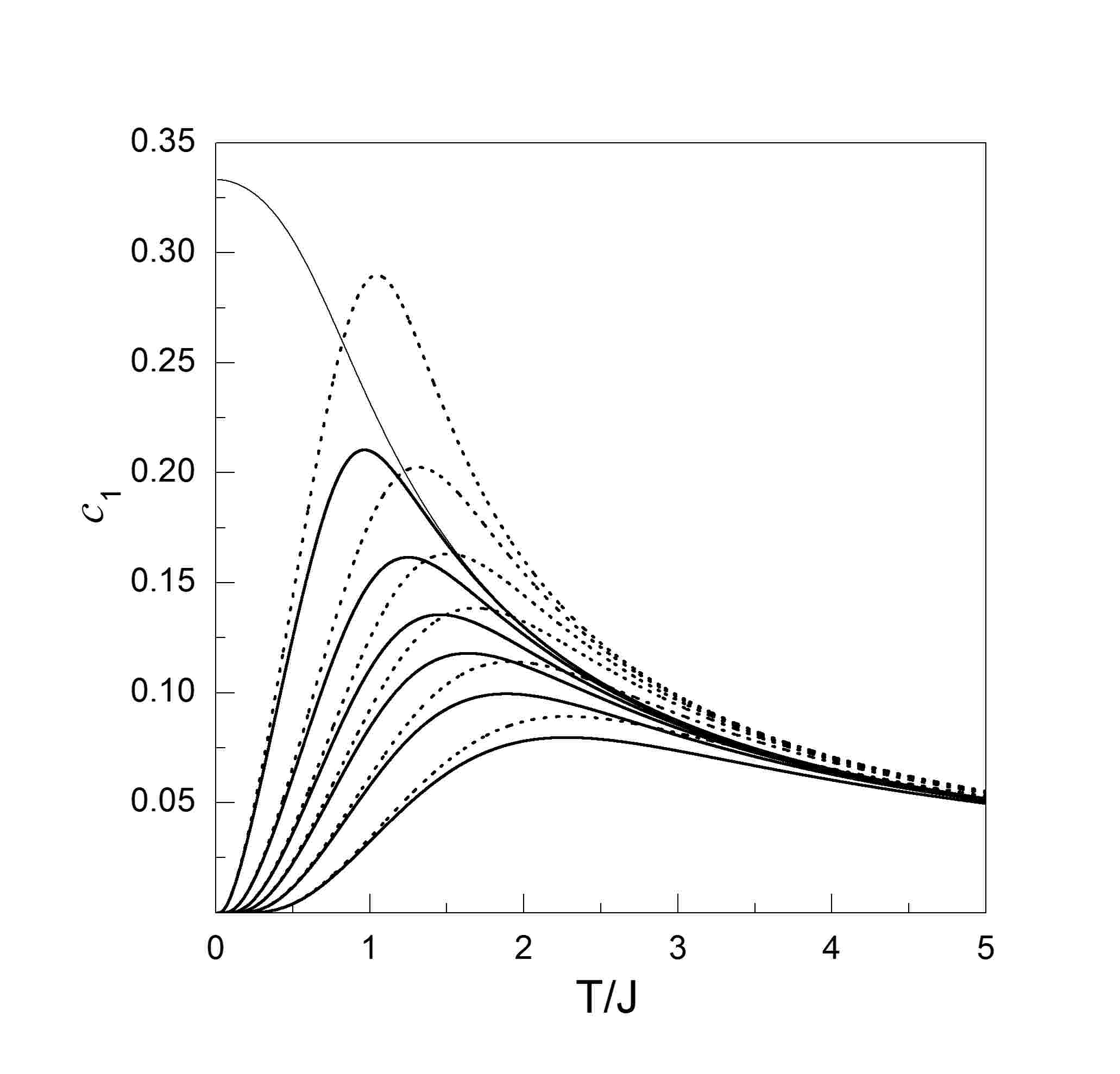}
\caption{\label{fig:epsart9}
Correlation functions $a_1$ (a) and
$c_1$ (b) for HFM on a triangular lattice at 
$h/J=$ 0.1, 0.3, 0.5, 0.7, 1.0, 1.5 (from left to right). 
The present theory (solid) and RPA (dotted). 
Thin lines correspond to $h=$0.}
\end{figure}
Temperature dependences of the correlation functions $a_1$ and $c_1$ 
at different $h/J$ are shown in Fig. \ref{fig:epsart9}.
\begin{figure}
\includegraphics[width=0.45\textwidth, trim=0 10 0 0]{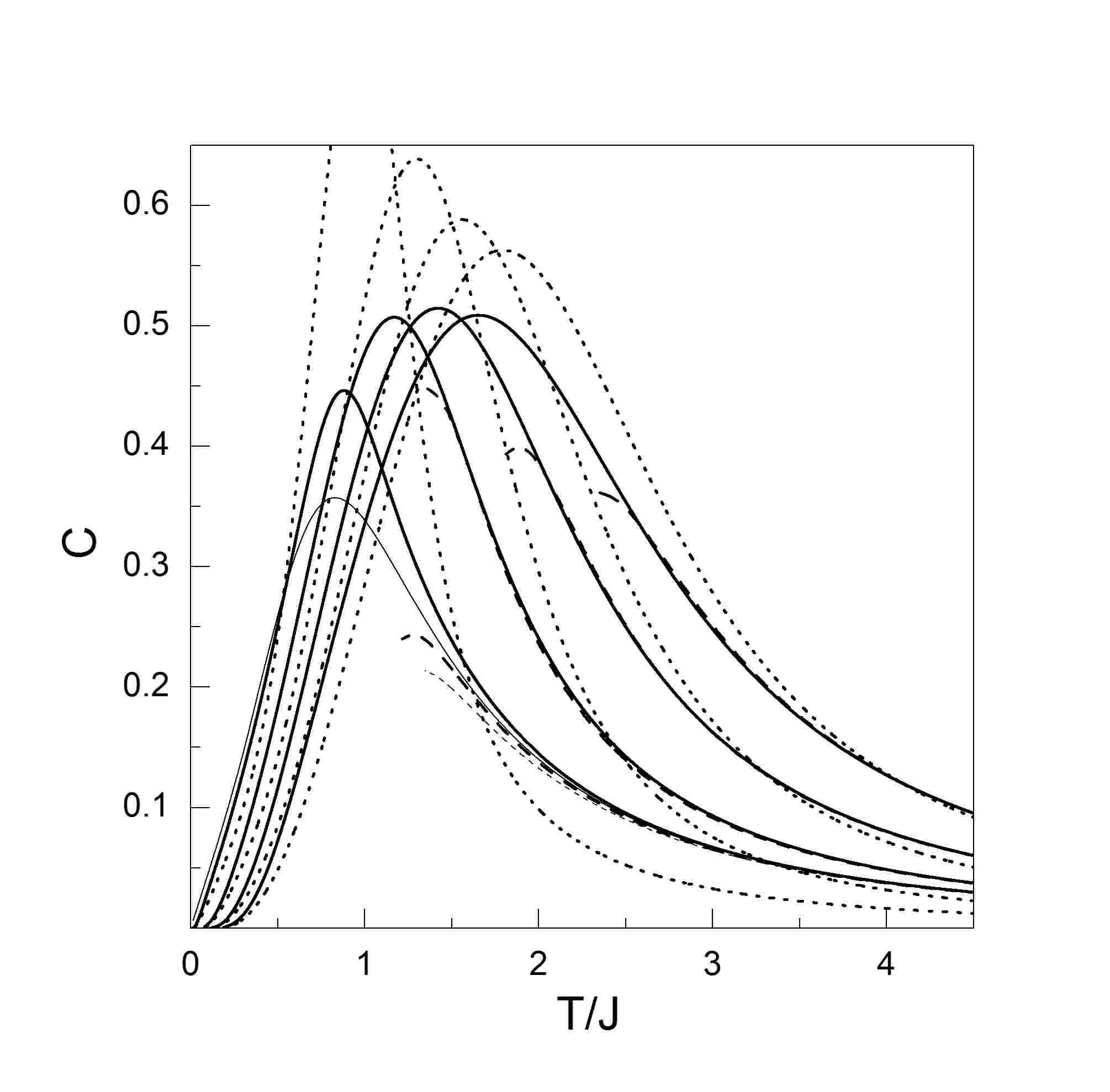}
\caption{\label{fig:epsart10}
Temperature dependences of the heat capacity for HFM on a triangular
lattice at $h/J=$ 0.1, 0.5, 1.0, 1.5 (from left to right). 
The present theory (solid), HTSE\cite{Baker} (dashed), and RPA (dotted).
Thin lines correspond to $h=$0.}
\end{figure}
Beginning with $T/J \sim 0.5$ the RPA results differ from ours sufficiently.
It is easy to verify that at $T\gg J$ the correlator $a_1$ calculated within 
RPA to the first approximation in $J/T$ is negative and equal to $-J/(4T)$. 
Thus, almost at all temperatures RPA fails to describe correctly the correlation 
functions and, hence, the energy and heat capacity.

\begin{figure}
\includegraphics[width=0.45\textwidth, trim=0 10 0 0]{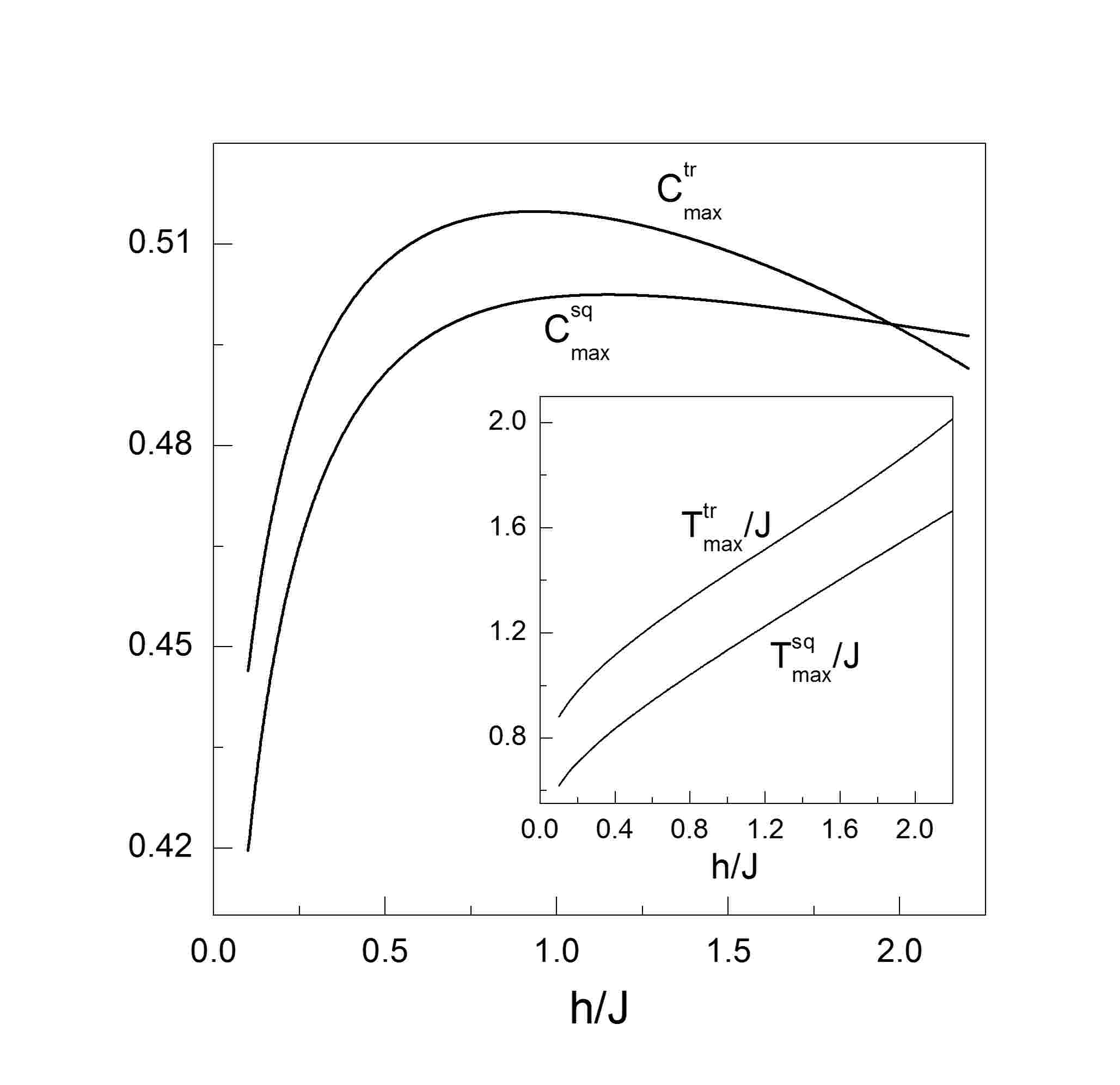}
\caption{\label{fig:epsart11}
Field dependences of the heights and positions of the heat capacity 
maxima for HFM on square and triangular lattices.}
\end{figure}

Fig. \ref{fig:epsart10} demonstrates the temperature dependences of 
the heat capacity $C(T)$ in comparison with HTSE \cite{Baker} and RPA. 
It is seen that our results are in good agreement with HTSE. With increase 
in field the position of the maximum in the curve $C(T)$ shifts to higher 
temperatures and its value $C_m^{tr}$ first increases rapidly and then decreases.
A similar behavior occurs for the heat capacity maximum  $C_m^{sq}$
on a square lattice. Maximum values $C_m^{sq}$ and $C_m^{tr}$ vs
field are illustrated in Fig. \ref{fig:epsart11}.
Field dependences of the maximum positions for the square and triangular 
lattices are shown in the inset. At $h/J\leq 1$ both $C_m^{tr}(h/J)$ and 
$C_m^{sq}(h/J)$ can be approximated by a function
\begin{equation}
C_m=\frac{ax}{b+x}, \qquad x=\frac{h}{J},\nonumber
\end{equation}
with
\begin{equation}
a=\left\{\begin{array}{l}
0.5136,\\
0.5254,\end{array}\right.   \qquad
b=\left\{\begin{array}{ll}
0.0239,& \quad{\rm square;}\\
0.0189,& \quad{\rm triangular.}
\end{array}\right.\nonumber
\end{equation}
At higher fields ($h/J\geq$ 1.4) the maximum values decrease linearly
\begin{equation}
C_m=Ax+B, \nonumber
\end{equation}
\begin{equation}
A=\left\{\begin{array}{l}
-0.00693,\\
-0.0248,
\end{array}\right.   
\qquad
B=\left\{\begin{array}{ll}
0.5118,& {\rm square;}\\
0.5468,& {\rm triangular.}
\end{array}\right.\nonumber
\end{equation}
Since $C_m^{sq}$ decreases slower than $C_m^{tr}$, the inequality $C_m^{tr}>C_m^{sq}$ 
valid for low fields changes into the opposite one at $h/J\geq 2$.

\begin{figure}
\includegraphics[width=0.45\textwidth, trim=0 10 0 0]{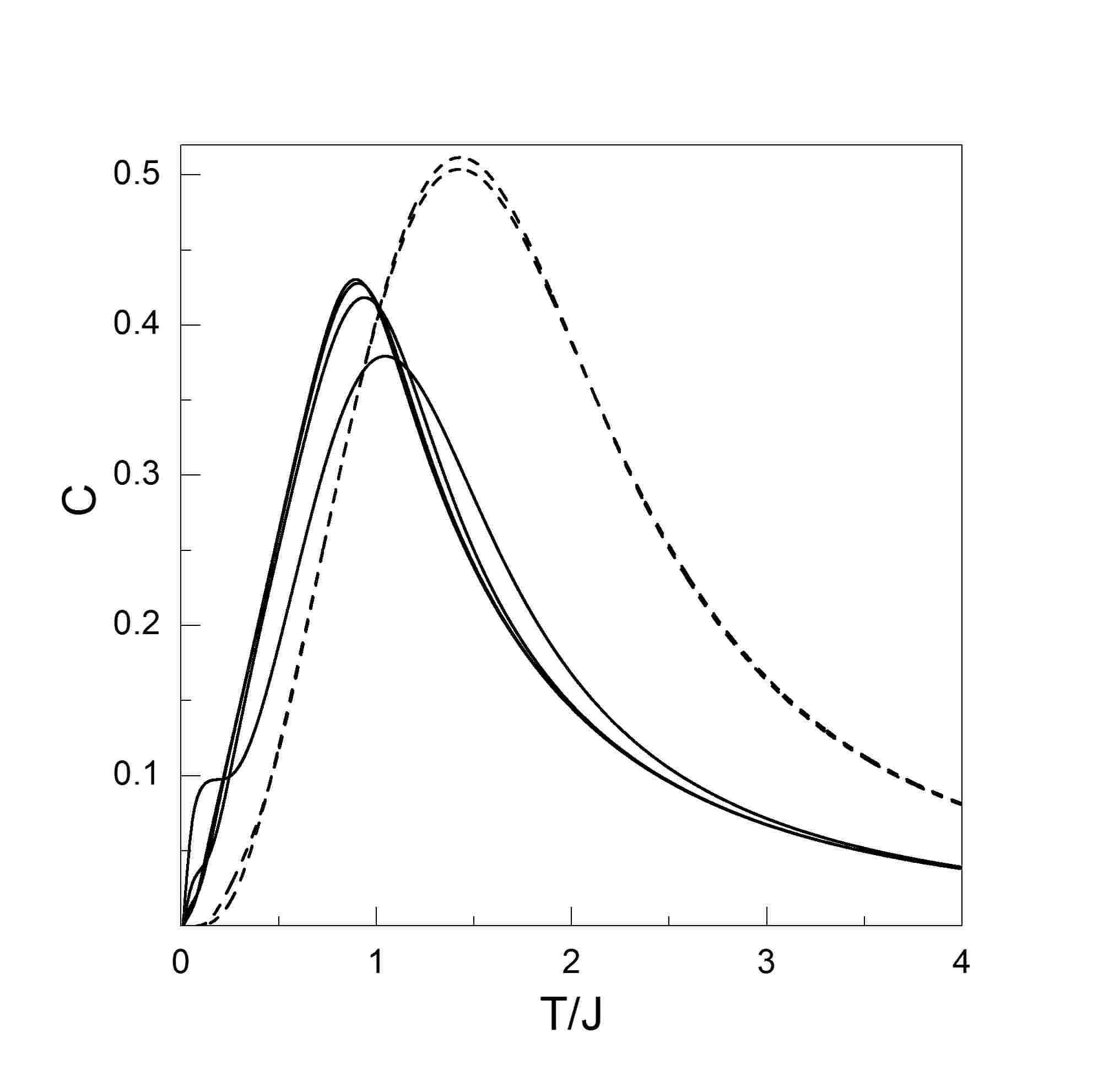}
\caption{\label{fig:epsart12}
Temperature dependences of the heat capacity
for HFM on a triangular lattice (from bottom to top):
$h/J=$0.1,\, $L=$4, 6, 8, 10 (solid) and $h/J=$1,\, $L=$4, 6 (dashed).}
\end{figure}
Let us consider now the dependence of the thermodynamic functions on
the cluster size $L\times L$. It is interesting to determine the linear 
size $L_0$ corresponding to the thermodynamic limit at a given magnetic 
field. This quantity is important, for example, on using such methods as 
Monte Carlo and exact diagonalization, when a knowledge of an optimal cluster 
size makes it possible to obtain the thermodynamic functions of the infinite
system within a reasonable volume of calculations. The dependence of the 
thermodynamic functions on $L$ is also of practical interest, because of the 
isle structure of $^3$He layers at some coverages. \cite{GR}

Fig. \ref{fig:epsart12} displays temperature dependences of the heat 
capacity at $h/J=$0.1 and 1 for different cluster sizes $L$ up to $L_0$. 
It is seen that with decrease in $L$ maximum in the curve $C(T)$ decreases 
and shifts to higher temperatures. At small $L$ and very low fields
a second maximum arises on the low temperature part of the heat capacity.
A similar additional maximum resulting from the finite size of the system 
was found by ED for the $4\times 4$ square lattice in Ref. \onlinecite{german1}.
This result is also reproduced by our calculations.

\begin{figure}
\includegraphics[width=0.45\textwidth, trim=0 10 0 0]{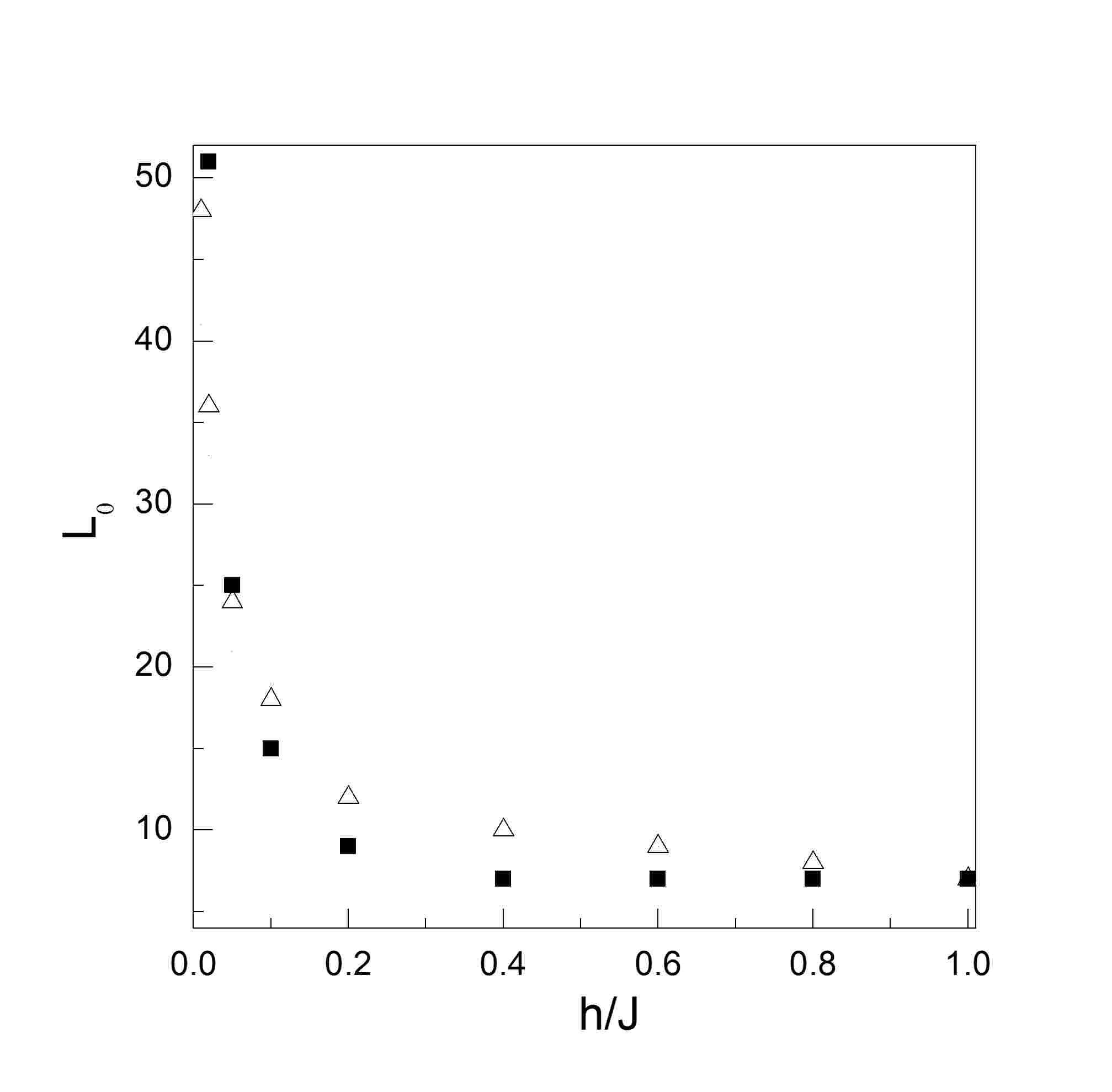}
\caption{\label{fig:epsart13}
Dependences $L_0(h/J)$ for triangular ($\triangle$) and
square ($\blacksquare$) lattices.}
\end{figure}
\begin{figure}[h]
\includegraphics[width=0.45\textwidth, trim=0 10 0 0]{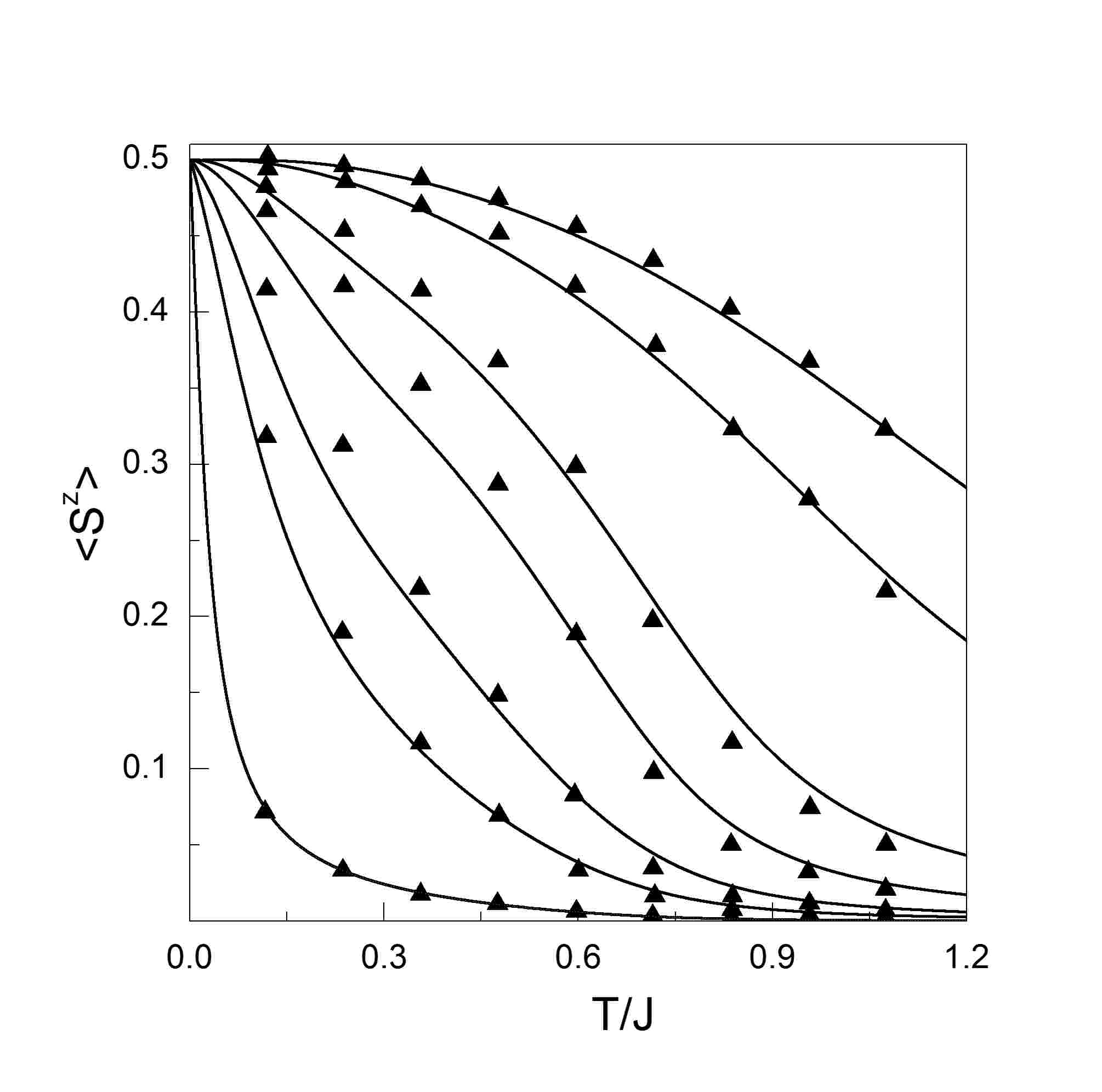}
\caption{\label{fig:epsart14}
Temperature dependences of the magnetization for $16\times 16$ 
triangular lattice HFM at $h/J=$0.429, 0.214, 0.0429, 0.0171, 
6.09 10$^{-3}$, 2.76 10$^{-3}$, 4.29 10$^{-4}$ (from top to bottom): 
present theory (solid) and QMC\cite{Kop} (symbols).}
\end{figure}
Fig. \ref{fig:epsart13} shows dependences $L_0(h/J)$ for the square 
and triangular lattices. At $h/J<0.2$ even small variation in field
leads to a sufficient change in $L_0$. As the field increases this
dependence weakens, so that beginning with $h/J\sim 0.2$ rather small-sized 
clusters are appropriate for the numerical simulations
of the real infinite systems.

Temperature dependences of the magnetization for a 16$\times$16 triangular
lattice together with the corresponding QMC data \cite{Kop} are shown
in Fig. \ref{fig:epsart14}. Our results agree well with QMC.

Now we check up the two criteria outlined in Sec. II, as applied to the
triangular lattice HFM. The limiting value of the entropy is equal to 
0.708 and 0.713 at  $h/J$=0.05 and 1, respectively, which slightly exceeds $\ln 2$.
At low and high temperatures the function $R(T,h)$ is close to zero as it was
for HFM on the chain and square lattice. At fixed field in the intermediate 
temperature region $R(T,h)$ has a maximum, whose height decreases as $h/J$ increases. 
The maximum value of $R$ is $\sim$0.094 at $h/J=0.05$, whereas at $h/J=1$ it does not 
exceed 0.046.

\section {SUMMARY}

The thermodynamics of the low dimensional spin-1/2 Heisenberg ferromagnets
in an external magnetic field is investigated within a second-order
two-time Green function formalism in the wide temperature and field 
range. The self-consistent set of equations for the correlation functions, 
vertex parameters and magnetization is obtained in the universal form 
appropriate for the description of low dimensional HFM on a chain, square 
and triangular lattices. The fundamental point of our consideration is the 
account of the correct analytical properties for the approximate transverse 
commutator Green function, from which the equation for the magnetization follows. 
This enables us to extend the range of adequate description for the HFM 
thermodynamics to lower fields as compared to the scheme proposed in Ref. 
\onlinecite{german1}.

The thermodynamics of a triangular lattice HFM in a magnetic field is
studied within a second-order Green function formalism for the first time.
The temperature dependences of the magnetization, susceptibility, correlation 
functions, and heat capacity at different values of the magnetic field are 
calculated and analyzed in detail. For square and triangular lattices the 
positions and heights of the heat capacity maxima vs field are obtained.
The dependences of the thermodynamic functions of the 2D HFM on the cluster 
size are investigated. For both types of lattices the cluster sizes corresponding 
to the thermodynamic limit are found as functions of field.

The temperature and field dependences for the thermodynamic functions
calculated within our scheme are in close agreement with the corresponding 
results obtained by Bethe ansatz, quantum Monte Carlo simulations, high temperature 
series expansion, and exact diagonalization. Thus, we can conclude that the scheme 
used in this paper provides a good quantitative description for the thermodynamics of 
the low dimensional HFM in an external magnetic field on the three considered types of 
lattices for infinite as well as for finite-sized systems.

\end{document}